\documentclass{article}
\usepackage{amsmath}
\usepackage{amssymb}
\usepackage[dvips]{graphicx}

\begin{document}

\title{INERTIAL MECHANISM: \\ DYNAMICAL MASS AS A  SOURCE
OF PARTICLE CREATION }

\author{A.V.~Filatov$^1$, A.V.~Prozorkevich$^1$, \\S.A.~Smolyansky$^1$ and V.D.~Toneev$^2$}

\date{\it $^1$Saratov State University, RU-410026, Saratov, Russia\\
$^2$Bogoliubov Laboratory for Theoretical Physics, \\Joint Institute
for Nuclear Research,\\ RU-141980, Dubna, Russia}

\maketitle

\begin{abstract}
A kinetic theory of vacuum particle creation under the action of
an inertial mechanism is constructed within a nonpertrubative
dynamical approach. At the semi-phenomenological level, the
inertial mechanism corresponds to quantum field theory with a
time-dependent mass. At the microscopic level, such a dependence
may be  caused by different reasons: The non-stationary Higgs
mechanism, the influence of a mean field or condensate, the
presence of the conformal multiplier in the scalar-tensor
gravitation theory {\em etc}. In what follows, a kinetic theory in
the collisionless approximation is developed for scalar, spinor
and massive vector fields in the framework of the oscillator
representation, which is an effective tool for transition to the
quasiparticle description and for derivation of non-Markovian
kinetic equations. Properties of these equations and relevant
observables (particle number and energy densities, pressure) are
studied. The developed theory is applied here to describe the
vacuum matter creation in conformal cosmological models and
discuss the problem of the observed number density of photons in
the cosmic microwave background radiation. As other example, the
self-consistent evolution of scalar fields with non-monotonic
self-interaction potentials  (the W~-~potential and Witten-Di
Vecchia - Veneziano model) is considered. In particular,
conditions for appearance of tachyonic modes and a problem of the
relevant definition of a vacuum state are considered.
\end{abstract}

PACS: 13.87.Ce \ 05.20.Dd \ 11.15.Tk \ 98.70.Vc \ 11.10.Lm

\section{Introduction}

The present  work is devoted to the construction of a kinetic
theory of  vacuum creation of particles with  time-dependent
masses. For brevity, this mechanism will be referred to as
inertial one. Microscopic foundations of a mass change may be
different. The Higgs mechanism leads to the most popular models of
such a class, when the corresponding mean fields are
time-dependent. General quantum field models with nonpolynomial
interactions may also be considered, where the separation of
nonstationary mean fields results in a time-dependent mass
\cite{PL05}. A well-known example of this kind  is the Witten --
Di Vecchia -- Veneziano model \cite{Veneziano,wit} in the
framework of which the mean-field concept was analyzed in
\cite{Blaschke}. The Nambu-Jona-Lasinio \cite{Volkov} and $\sigma$
\cite{Koch} models are other examples, where the meson masses are
defined by evolution of a quark condensate to be described at the
hydrodynamic \cite{Baier} or kinetic \cite{Florkowski} level. The
particle mass may depend on many-particle interactions in hot and
dense non-stationary matter \cite{Pisarski,FHL,Cooper}. A general
basis for a rather slowly-varying time dependence of the effective
mass can be obtained within the Green function method
\cite{Muller,Maino}. The field dependence of the mass is a general
factor determining the time evolution in all these cases (F-class
models). The conformal invariance of the scalar-tensor
gravitational theory provides a time dependence of the particle
mass by means of the conformal multiplier
\cite{FM,Prokopec,DB,Perv02}. The mass can be changed also due to
the parametrization stipulated by additional space dimensions
\cite{Slavnov}. Such theories should be referred to as the other
class (C-class). In the F-class theories, the vacuum particle
creation admits a well-known interpretation based on the
simplified vacuum tunnelling model in an external field
\cite{Sauter1,HE,Schwinger}. A similar interpretation of the
C-class models is difficult. At the phenomenological level,
however, both classes have a uniform mathematical description as
it will be shown in Sects. \ref{SF}, \ref{secf} and \ref{MVB},
respectively, for the scalar, fermion and massive vector boson in
quantum field theories (QFT's).

The first consideration of the vacuum creation of particles with
the variable mass was proposed apparently in \cite{Andreev02} as a
possible variant for describing a quantum system response to the
time variation of system parameters \cite{Andreev96}. Using the
Bogoliubov transformation method, residual momentum distributions
for  fermions,   and pair correlators were found for the cases of
step-like and smooth variations of the fermion mass (Sect.
\ref{secf4}).

In the present work, the kinetic  theory will be based on the
oscillator representation (OR) \cite{OR,Perv99}, which is the most
economical method for a nonpertrubative description (as compared
to the Bogoliubov method of  canonical transformation \cite{Grib}
or other accurate approaches to the problem \cite{Prokopec, bgr,
Krekora}) of the vacuum particle creation under the action of
time-dependent strong fields. This approach leads directly to the
quasiparticle representation (QPR) with diagonal operator forms in
the momentum space for the set of dynamical variables. It allows
one to get easily  the Heisenberg-type equations of motion for
creation and annihilation operators. An important feature of the
time-dependent Fock representation is the necessary consistence of
commutation (anti-commutation) relations with the equations of
motion. Otherwise, this circumstance can bring to the
non-canonical quantization rules (an example will be considered in
Sect. \ref{MVB1}).

In terms of the OR  it is possible to immediately derive the
corresponding kinetic equations (KE) by the well known method
\cite{Schmidt}. Some particular results are published in
\cite{PL05,Blaschke2}. The kinetic theory for scalar, spinor, and
massive vector fields is constructed in Sects. \ref{SF},
\ref{secf} and \ref{MVB}, respectively. The main attention is paid
to the particle creation in conformal cosmological models
\cite{ZZ97,BPZZ06,BBPP02} (Sec. \ref{UNI}). It is shown that the
choice of the equation of state (EoS) of the Universe allows one
to obtain, in principle, the observed number density of matter
participants and photons and, possibly, dark matter. The basic
problem here is the description of vacuum particle creation which
should be consistent with EoS but it is beyond the present
article.

Finally, in the Sect. \ref{meta}  other class of scalar QFT
systems is considered with non-monotonic self-interaction
potentials to apply the decomposition of the field amplitude into
the quasiclassical space-homogeneous time-dependent background
field and the fluctuation part. In this case, the particle mass is
defined by intensity of a quasiclassical field. As an example,
self-interaction potentials of the simplest polynomial type and
that for a nontrivial case (the pseudoscalar sector of the Witten
-- Di Vecchia -- Veneziano model) are analyzed. It is shown, that
the relevant definition of vacuum states allows one to avoid of
the tachyonic mode beginning. The main purpose of this review is
to summarize  all known relevant results on the inertial mechanism
of the vacuum particle production and to call attention to
unsolved problems which are shortly listed in Sect. \ref{SUM}.

We use the metric $g^{\mu\nu}=diag(1,-1,-1,-1)$ and natural units
$\hbar=c=1$.

\section{Scalar field \label{SF}}

\subsection{Oscillator and quasiparticle representations \label{2.1}}

Let us start our consideration with  the simplest case of the real
scalar field with the time-dependent mass $m(t)$, whose equation
of motion is
\begin{equation}
\label{gen_eq_of_mot} [\partial_\mu\partial^\mu +
m^2(t)]\varphi(x) = 0.
\end{equation}
The corresponding Lagrange function is given as
\begin{equation}\label{lagr}
    L=\frac12\partial_\mu \varphi\, \partial^\mu  \varphi -\frac12 m^2(t)
    \varphi^2.
\end{equation}

In the considered case, the system is space-homogeneous and
nonstationary. Therefore, the transition to the Fock space can be
realized on the basis functions $\varphi(\mathbf{x}) \sim exp(\pm
i \mathbf{p x})$, and creation and annihilation operators become
time-dependent. The assumption about the space homogeneity allows
one to look for solution of Eq. (\ref{gen_eq_of_mot}) in discrete
momentum space in the following form:
\begin{equation} \label{disc_mom_sp} \varphi(\mathbf{x},t) =
\frac{1}{\sqrt{V}}\sum_{\mathbf{p}} e^{i \mathbf{p} \mathbf{x}}
\varphi(\mathbf{p}, t)~,
\end{equation}
where $V = L^3$ and $p_i = (2\pi / L)n_i$  at that the integers
$n_i, \ (i=1,2,3) $ run from $-\infty$ to $+\infty$. The
thermodynamic limit can be covered in the resulting equations.
Then the oscillator-type equation of motion follows from Eqs.
(\ref{gen_eq_of_mot}) and decomposition (\ref{disc_mom_sp}) as
\begin{equation}  \label{osc_eq_of_mot} \ddot\varphi^{(\pm)}(\mathbf{p}, t) +
 \omega^2(\mathbf{p},
t)\varphi^{(\pm)}(\mathbf{p}, t) = 0
\end{equation}
with
\begin{equation}
\label{tm_freq} \omega^2(\mathbf{p}, t) = {m^2(t) +
\mathbf{p}^2}~.
\end{equation}
The symbols $(\pm)$  correspond to the positive and negative
frequency solutions of Eq. (\ref{osc_eq_of_mot}) defined by its
free asymptotics in the infinite past (future) \cite{Grib},
\begin{equation} \label{free_assympt} \varphi^{(\pm)}(\mathbf{p},
t \to \mp\infty) \ \sim  e^{\pm i \omega_{\mp}t},
\end{equation}
where $\omega_{\mp}=\sqrt{m_{\mp}^2+\mathbf{p}^2}$ are defined by
asymptotics of the mass
\begin{equation}
\label{mass_assympt} m_{\mp} = \lim_{t \to \mp\infty} m(t).
\end{equation}
The asymptotics (\ref{free_assympt}) corresponds to the
in(out)-states and is necessary for definition of in(out)-vacuum.
This requirement, however, can be broken in cosmology
\cite{Birrell}. We suppose here that such asymptotics exists and
the relevant vacuum states will be denoted by $|0\rangle$ without
indices "in"\, or "out"\,, that is evident from the context. In
the considered class of problems, the classification of states in
the frequency sign turns out to be impossible for an arbitrary
time moment. According to a general analysis \cite{Fradkin}, this
leads to instability of a vacuum state during the action period of
the external fields and to the vacuum particle creation. In this
case, it is possible to consider quasiparticle excitations during
the system evolution ($\dot{m}(t) \neq 0$) and describe their
creation and annihilation in the vacuum state $|0\rangle$. When
the external field action is completed,  residual particles of
some finite density remain in the out-state. However, it is
necessary to emphasize that these particles are defined in respect
of the in-vacuum  state \cite{Grib}.

In the general case, S-matrix does not exist in the considered
formalism.  Its role in description of the system with unstable
vacuum is performed by other mathematical objects: The operator of
the canonical Bogoliubov transformation \cite{Grib}, distribution
functions in the kinetic approach \cite{OR,Schmidt,
Blaschke2,Schmidt2,Schm4,Vinnik} or some set of correlation
functions \cite{quark,Casimir06} {\em etc}.

The conception of "quasiparticle" plays the central role in the
QFT with strong time-dependent quasiclassical external fields
\cite{Grib, Fradkin}. Under the considered conditions, this
approach is the shortest realization of the quasiparticle concept
within the standard Fock representation of the QFT, where the
external field can be taken into account nonperturbatively. Thus,
the QPR corresponds to a possibility of writing down the set of
commutative operators of physical (observable) quantities (the
complete QPR) in the diagonal form in an arbitrary time. It is
naturally related to the question whether operators have the
quadratic form in the Fock representation. Hence, the interaction
between the field constituents and self-interaction is not taken
into account. It corresponds to a nondissipative approximation in
the kinetic theory \cite{Kadanoff}. An alternative definition of
the quasiparticle was given in \cite{Perv99} for constrained
systems.

The transition to the QPR  can be realized in different ways.  The
traditional method is based on the time-dependent canonical
Bogoliubov transformation \cite{Grib}. The alternative approach
uses the oscillator ("holomorphic") representation, which leads
directly to the QRP \cite{OR}. In the considered case the
transition to the OR is made by substituting $m_{\pm} \to m(t)$
into the dispersion law for the  free field and postulating a
following decompositions:
\begin{align} \label{8} \varphi(x) =
\frac{1}{\sqrt{2V}}\sum_{\mathbf{p}}\frac{1}{\sqrt{\omega(\mathbf{p},
t)}} \biggl\{a(\mathbf{p}, t)\, e^{i\mathbf{p} \mathbf{x}}  +
a^{\dagger}(\mathbf{p}, t)\, e^{-i\mathbf{p} \mathbf{x}} \biggr\} , \nonumber \\
\pi(x) = -
\frac{i}{\sqrt{2V}}\sum_{\mathbf{p}}\sqrt{{\omega(\mathbf{p},
t)}}  \left\{a(\mathbf{p}, t)\, e^{i\mathbf{p} \mathbf{x}} -
 a^{\dagger}(\mathbf{p}, t)\, e^{-i\mathbf{p} \mathbf{x}} \right\},
\end{align}
where $\pi(x)$ is the generalized momentum,
$a^{\dagger}(\mathbf{p}, t)$ and $a(\mathbf{p}, t)$ are the
creation and annihilation operators of particles with the momentum
$\mathbf{p}$ at the time moment $t$. The in-vacuum state is
defined as
\begin{equation}\label{9}
a(\mathbf{p}, t\to -\infty)|0> \,=\, 0,   \qquad\qquad <0|0> = 1.
\end{equation}
The canonical commutation relation
\begin{equation} \label{can_commutator} \left[ \varphi(x), \pi(x')
\right]_{t=t'} = i \delta(\mathbf{x} - \mathbf{x'})
\end{equation}
together with the decomposition (\ref{8}) provides the standard
commutation relation for time-dependent creation and annihilation
operators
\begin{equation} \label{st_eq_of_mot} \left[ a(\mathbf{p}, t),
a^{\dagger}(\mathbf{p'}, t) \right] = \delta_{\mathbf{pp'}}~.
\end{equation}

The substitution of  the decompositions (\ref{8})
into the Hamiltonian
\begin{equation}
\label{Hamiltonian} H(t) = \frac{1}{2}\int d^3x\left\{ \pi^2(x) +
[\nabla\varphi(x)]^2 + m^2(t)\varphi^2(x)\right\}
\end{equation}
leads immediately to a diagonal form which corresponds to the QPR
\begin{equation} H(t) = \sum\limits_{\mathbf{p}} \omega(\mathbf{p}, t) \left\{
a^{\dagger}(\mathbf{p}, t)a(\mathbf{p}, t)  + \frac{1}{2} \right\}
.
\end{equation}
In the considered case the vacuum energy of zero oscillations
("Zitterbewegung"\,) depends on time.

Equations of motion for the operators $a, a^{\dagger}$ can be
obtained now from  the minimal action principle \cite{OR} or from
the Hamiltonian equations
\begin{equation}\label{hameq}
    \dot{\varphi} = \frac{\delta H}{\delta \pi}=\pi,\qquad\qquad
    \dot{\pi} = -\frac{\delta H}{\delta \varphi}= \triangle \varphi
    -m^2(t) \varphi~.
\end{equation}
Here and below we use the notation $(\dot{\ }) = d/dt(\ )$.
Combining Eqs. (\ref{8}) and (\ref{hameq}), we get
\begin{align} \dot{a}(\mathbf{p}, t)
= \frac{1}{2} \Delta(\mathbf{p}, t)\, a^{\dagger}(-\mathbf{p}, t)
- i
\omega(\mathbf{p}, t)\, a(\mathbf{p}, t), \nonumber \\
\dot{a}^{\dagger}(\mathbf{p}, t) = \frac{1}{2} \Delta(\mathbf{p},
t) \, a(-\mathbf{p}, t)  + i \omega(\mathbf{p}, t)\,
a^{\dagger}(\mathbf{p}, t),\label{eq_of_mot}
\end{align}
where
\begin{equation} \label{Delta} \Delta(\mathbf{p}, t) = \frac{\dot \omega(\mathbf{p},
t)}{\omega(\mathbf{p}, t)} =  \frac{m(t)\, \dot
m(t)}{\omega^2(\mathbf{p}, t)} \end{equation}
 is the factor defining the mixing of  states with positive and
negative energies. This equation obviously is consistent with the
commutation relations (\ref{st_eq_of_mot}).

Equations of motion (\ref{eq_of_mot}) can be rewritten as the
Heisenberg-type  equation, e.g.,
\begin{equation} \label{Hei_eq_of_mot} \dot{a}(\mathbf{p}, t) = \frac{1}{2}
\Delta(\mathbf{p}, t) a^{\dagger}(-\mathbf{p}, t)  + i\left[ H(t),
a(\mathbf{p}, t) \right]~.
\end{equation}
In the instantaneous QPR these equations serve as a basis  for a
nonperturbative derivation of the KE describing  scalar particle
creation and annihilation processes within the inertial mechanism.

\subsection{Kinetic equation \label{KinEq}} 
The key object of the kinetic theory is the quasiparticle
distribution function which for the space-homogeneous case is
\begin{equation}
\label{distr_func} f(\mathbf{p}, t) = \langle 0|
a^{\dagger}(\mathbf{p}, t) a(\mathbf{p}, t)|0 \rangle,
\end{equation}
where $|0\rangle = |0\rangle_{in}$ is the initial ($t\to -\infty$)
vacuum state. Differentiating the distribution function
(\ref{distr_func}) with respect to time and  using (\ref{eq_of_mot}) we get
\begin{equation} \label{diff_eq_of_mot} \dot{f}(\mathbf{p}, t) = 
\Delta(\mathbf{p}, t) \mathrm{Re}\{f^{(+)}(\mathbf{p}, t)\} .
\end{equation}
Here the auxiliary correlation function is introduced
\begin{align}\label{distr_eq_of_mot}
f^{(+)}(\mathbf{p}, t) &= \langle 0| a^{\dagger}(\mathbf{p}, t)
a^{\dagger}(-\mathbf{p}, t) |0 \rangle .
\end{align}
This function provides a coherent connection between the states
with positive and negative energies (so-called entangled states
\cite{Krekora}).

The equation of motion for $ f^{(+)}(\mathbf{p}, t)$ can be
obtained by analogy  with  equation (\ref{diff_eq_of_mot}). We
present it here in the integral form as
\begin{equation} \label{Int_distr_func} f^{(+)}(\mathbf{p}, t) =
\frac12 \int\limits_{t_0}^t dt' \Delta(\mathbf{p}, t') \left[ 1 +
2f(\mathbf{p}, t')  \right] e^{2i \theta(\mathbf{p}; t, t')},
\end{equation}
where the initial condition $f^{(+)}(\mathbf{p}, t_0) = 0$ was
used.  This  condition corresponds to the  initial condition for
the distribution  function   $f(\mathbf{p},  t_0)  = 0$ and is a
direct consequence of the definition (\ref{9}). Eventually, the
dynamical phase in Eq.(\ref{Int_distr_func}) is equal to
\begin{equation}
\theta(\mathbf{p}; t, t') = \int\limits_{t'}^t d\tau
\omega(\mathbf{p}, \tau).
\label{theta}
\end{equation}

The substitution of Eqs.(\ref{Int_distr_func}) in
Eq.(\ref{diff_eq_of_mot})  leads us to the resulting KE  written
in the thermodynamical  limit $V \to \infty$ at the fixed particle
density
\begin{equation}
\label{I_KE} \dot{f}(\mathbf{p}, t) =  
\frac{1}{2} \Delta(\mathbf{p}, t) \int\limits_{t_0}^t dt'
\Delta(\mathbf{p'}, t)  \left[ 1 + 2f(\mathbf{p}, t') \right] \cos
\left[ 2\theta(\mathbf{p}; t, t') \right].
\end{equation}
The source term in the r.h.s. of Eq. (\ref{I_KE}) describes a
variation of the particle number with the given momentum due to
vacuum creation and annihilation processes for the inertial
mechanism, $\Delta(\mathbf{p}, t)$ is defined by Eq.(\ref{Delta}).
The non-Markovian  KE (\ref{I_KE}) has the structure as that for
the Schwinger mechanism of pair creation in an electric field
\cite{Schmidt}. This equation was investigated in detail for
description of the pre-equilibrium evolution of quark-gluon plasma
created in collisions of ultrarelativistic heavy ions
\cite{Schmidt2,Schm4}. The case of the scalar QED was considered
in \cite{OR} for an electric field of arbitrary polarization.

As it follows from Eq.(\ref{Delta}), in the framework of the
inertial mechanism the particle production rate is defined by the
rate of the mass change
\begin{equation}\label{mch}
\xi (t) = \frac{1}{m(t)} \frac{dm(t)}{d(m_{0} t)},
\end{equation}
where $m_0$ is the characteristic mass to fix the time scale ({\em
e.g.}, $m_0 = m_-$).

Let us note that the KE (\ref{I_KE}) is valid under two basic
assumptions: a) there are no particles (or antiparticles)  in the
in-state; b) a collisionless approximation is applicable ({\em
i.e.} the corresponding dissipative processes are not taken into
consideration).

In the low-density approximation $f(\mathbf{p},t) \ll 1 $ the KE
(\ref{I_KE}) results in the following solution \cite{Schmidt2}:
\begin{equation}\label{star}
f(\mathbf{p},t) = \frac{1}{4}  \left| \int\limits_{t_0}^{t} dt'
\Delta  (\mathbf{p},t') \exp{[2i\theta
(\mathbf{p};t,t')]}\right|^2 \geq 0.
\end{equation}
The KE (\ref{I_KE}) can be transformed to  linear equations of the
non-Hamiltonian dynamical system   with zero initial conditions
\begin{eqnarray}
\label{ODE_SYS} \dot{f} = \frac{1}{2} \Delta u, \qquad \dot{u} =
\Delta(1 + 2f) - 2\omega v,\qquad \dot{v} = 2 \omega u ,
\end{eqnarray}
which is convenient for numerical analysis. This equation system
has the first integral
\begin{equation}\label{f_int}
    (1+2f)^2 -u^2 -v^2=1,
\end{equation}
according to which the phase trajectories are located on the
two-cavity hyperboloid with top coordinates $f=u=v=0$ (physical
branch) and $f=-1, u=v=0$ (nonphysical one). If the function $f$
is excluded from Eqs.  (\ref{ODE_SYS}), we obtain the non-linear
two-dimensional dynamical system with
\begin{eqnarray}\label{two}
   \dot u&=&\Delta\sqrt{1+u^2+v^2}-2\omega v,\nonumber\\
    \dot v&=&2\omega u^{\phantom{\mathstrut}}.
 \end{eqnarray}
The functions $f(\mathbf{p},t)$ and $u(\mathbf{p},t)$ have a
certain physical meaning (the last function describes  vacuum
polarization effects, see Sect. \ref{SF3}) below and are
invariants with respect to the time inversion $t \to -t$ while the
auxiliary function $v(\mathbf{p},t)$ and factor (\ref{Delta})
change their signs. Thus, the KE (\ref{I_KE}) is invariant at the
time inversion.

 The presented formalism of vacuum particle creation is specific
for kinetic theory and allows one natural generalization to the
case of interacting fields that leads to introduction of
corresponding collision integral of a non-Markovian type
\cite{CI}. This approach is close to the modern method expounded
in the book~\cite{Grib} where the time-dependent Bogoliubov
transformation is used. The same method was used in pioneer works
 \cite{TCh68,BTCh68,Parker} (see also \cite{Birrell}). Some
modification of the formalism \cite{Grib} (the $r,\bar{\theta}$
representation) was developed in \cite{Perv99} and then used
widely ({\it e.q.}, in \cite{BPZZ06} and references cited there).
The correspondence between the $r,\bar{\theta}$ representation and
our approach (as well as that used in the book \cite{Grib}) may be
easily established.

\subsection{Observable and regularization \label{SF3}}

The KE (\ref{I_KE}) describes the vacuum quasiparticle excitations
rising at an external force ($\dot{m}(t) \neq 0$, in the
considered case). When  this action is switched off, there is
still some remaining density of real (residual) particles  and
antiparticles. In the absence of any interaction between the
system constituents, the real particles are "on-shell" ones and
have the free-particle dispersion law $\omega _{\pm}$  with the
mass $m_{\pm}$ (\ref{mass_assympt}), while quasiparticles are
"off-shell"  with the dispersion law (\ref{tm_freq}).  Within the
Green function method \cite{Maino}, one can say that the
time-dependent dispersion law like (\ref{tm_freq}) corresponds to
the t-parametric mass shell surface of slowly time-dependent
$m(t)$, {\em i.e.} $m(T\pm \tau /2)\approx m(T)$, where $T$ and
$\tau$ correspond to slow and fast time scales. This case is not
of interest for the considered problem. Thus, the dispersion law
(\ref{tm_freq}) does not belong to the mass shell surface. In the
general case, the on-shell condition
\begin{equation}\label{on_shell}
\left|\frac{m(t) - m_0}{m_0} \right| \ll {1}
\end{equation}
($m_0 = m_{\pm}$) is not connected directly with the condition of
efficiency of the vacuum particle creation, $\xi (t) \lesssim 1$,
where $\xi (t)$ is defined by Eq.(\ref{mch}). On the contrary, the
presence of high frequencies in the function $m(t)$ is necessary
for vacuum creation and does not contradict the on-shell condition
(\ref{on_shell}). In principle, the KEs of such a type are
designed for the description of evolution of both real particles
and quasiparticles. In particular, the distribution function of
residual particles is $f_{out} (\mathbf{p}) =
\lim\limits_{t\to\infty} f(\mathbf{p},t)$. This simple formula for
$f_{out} (\mathbf{p})$ follows from  Eq. (\ref{star}) in the
low-density approximation. However, the presence of the fast
oscillated multiplier in the source term in the r.h.s. of the KE
(\ref{I_KE}) leads to a large amount of numerical calculations
which make impossible the study of the system evolution for rather
large times  after the switching on external forces. The
corresponding large scaling methods of calculations based on the
KE (\ref{I_KE}) have not been worked out at present.  Some
properties of a residual particle-antiparticle plasma due to a
limited pulse of the external field action can be estmated by the
imaginary time method \cite{Popov}.

The distribution function is the key quantity of the system. The
density of observable variables is some integral in the momentum
space containing the distribution function and auxiliary functions
$u(\mathbf{p}, t)$, which describe the effects of vacuum
polarization. The simplest variable of such a type is the
density of quasiparticles. In the thermodynamic limit $L\to\infty$ we have
\begin{equation}
n_{tot}(t) = \int [dp] f(\mathbf{p}, t),\label{dens}
\end{equation}
where $[dp] = (2\pi)^{-3} d^3p$. To
proceed to the thermodynamical limit,  the rule
\begin{equation}\label{term-lim}
    \frac{1}{L^3} \sum\limits_{\mathbf{p}} \to \int [dp]
\end{equation}
is used here and below.

Other important characteristics of the system are the energy
density $\varepsilon$ and pressure $P$, which can be obtained as
the average value of the energy-momentum tensor corresponding to
the Lagrangian density (\ref{lagr}),
\begin{equation}\label{tik}
    T_{\mu\nu} = \partial_\mu \varphi \partial_\nu \varphi -g_{\mu\nu}L.
\end{equation}
As the result we have \cite{Grib}
\begin{align}\label{t00}
\varepsilon &= \, \langle 0|T_{00}|0 \rangle \, =  \int [dp]\,
\omega f,\\ 3 P & = \varepsilon - \int [dp] \biggl[
\frac{m^2}{\omega}\bigl(f + \frac12 u \bigr) + \omega u \biggr].
\label{presss}
\end{align}
The last two terms in integrand (\ref{presss}) represent the
contribution of vacuum polarization.

Finally, the entropy density can be introduced
\begin{equation}\label{entropy}
 S(t) = -\int [dp] \bigl[f\ln{f} -(1+f)\ln{(1+f)} \bigr].
\end{equation}
It is not conserved ($\dot {S} (t) \neq 0$) even in the considered
non-dissipative approximation because the system is open (the mass
change is defined by external causes).

A direct proof of the convergence of  integrals (\ref{dens}),
(\ref{t00}), (\ref{presss}), (\ref{entropy}) is complicated
because of the absence of an explicit form for functions
$f(\bf{p}, t)$ and $u(\bf{p}, t)$. Therefore, one usually uses the
method of asymptotic expansions in power series of the inverse
momentum $p^{-N}$ (N-wave regularization technique) \cite{nw}
(another approach rests on the WKB approximation \cite{Casher}).
Our present consideration is based on the explicit asymptotic
solutions of the system (\ref{ODE_SYS}) for $|\mathbf{p}| \gg m$.
The integral (\ref{dens}) is assumed to be convergent at any time
moment. Then the function $f(\mathbf{p}, t)$ should decrease at
$\mathbf{p} \to \infty$ and hence $f(\mathbf{p}, t) \ll 1$ in this
region. This inequality corresponds to the low-density
approximation (\ref{star}), where the KE solution can be written
in the explicit form as
\begin{equation}
f^{\infty}(\mathbf{p}, t) = \frac{1}{4{p}^4} \biggl|
\int\limits_{t_0}^t dt' m(t') \dot{m}(t') \exp{[ 2ip\,(t -
t')]}\biggr|^2, \label{32}
\end{equation}
$p = |\mathbf{p}| \to \infty$ because
$\Delta^{\infty}(\mathbf{p},t) = m(t) \dot{m} (t) / p^2$ in
accordance with Eq. (\ref{Delta}). These solutions are consistent
with the integral of motion (\ref{f_int}). Thus, indeed asymptotic
solutions are some quickly oscillating functions (this fact was
first noted  in \cite{Blaschke2} for the case of massive vector
bosons, see Sect. \ref{EOS}). Such behavior matches with the
quasiparticle interpretation of vacuum excitations by the inertial
mechanism. The real (observed) particles are the result of the
evolution by the moment when $\dot{m}=0$ and the out-vacuum state
is realized. The asymptotics (\ref{32}) may be influenced by other
(non-inertial) mechanisms of vacuum particle creation ({\em e.g.},
in the case of harmonic "laser" electric field
\cite{laser,laser1,laser_phys_rev_lett_96}).

The asymptotics of the integral (\ref{32}) can be obtained by the
stationary phase method \cite{Fedor}
\begin{equation}
\label{charac_integr} \int\limits_{t_0}^t dt'\ m(t') \dot{m}(t')
e^{2ip(t-t')} = \ \frac{m(t)\dot{m}(t)}{ip} + O(p^{-2}),
\end{equation}
if $\dot{m}(t_0) = 0$. Using Eqs. (\ref{32}) and (\ref{ODE_SYS})
we get the leading contributions
\begin{eqnarray}
\label{lead_terms} f^{(6)}(\mathbf{p}, t) &=&
\left[\frac{m(t)\dot{m}(t)}{2 p^3}\right]^2, \nonumber \\[5pt] u^{(4)}(\mathbf{p},
t) &=& \frac{1}{p^4} \biggl[\dot{m}^2(t) + m(t)\ddot{m}(t)\biggr],
\end{eqnarray}
where the upper indices show the inverse momentum degree for the
corresponding leading terms (we are not  interested in the
asymptote of the function $v(\mathbf{p},t)$, which plays some
auxiliary role only). Relations (\ref{lead_terms}) are identical
to the results of application  of the N-wave regularization method
to Eqs.(\ref{ODE_SYS}) \cite{nw}.

Now one can conclude that the integral (\ref{t00}) is convergent
but the last integral term in Eq. (\ref{presss}) needs a
regularization. The regularizing procedure of the Pauli-Villars
type is based on the subtraction of appropriate counterterms in
integrals (\ref{dens}), (\ref{t00}), (\ref{presss}),
(\ref{entropy}). These counterterms can be obtained by the
substitution $p^2 \rightarrow p^2 + M^2$ into the denominator of
asymptotics (\ref{lead_terms}),
\begin{equation}
 f_R = f - f_M,   \qquad    u_R = u - u_M .
 \end{equation}
If the regularizing mass $M \gg m(t)$ can be chosen  rather large,
$M \gg \Lambda$ ($\Lambda$ is "the computer cut-off parameter"),
the influence of counterterms  on the results of numerical
calculations is negligible.

The numerical investigation of the KE (\ref{I_KE}) and observable
densities (\ref{dens}), (\ref{t00})-(\ref{entropy}) will be
presented in Sect. \ref{secf4}.

\section{Fermion field \label{secf}}

\subsection{Quasiparticle representation \label{secf1}}

The material of this subsection is based on papers
\cite{49,SPIE05,Skokov}.

Equations of motion for fermion fields with the variable mass are
\begin{eqnarray}
\label{eq_of_mot_w_alt_mass} [i \gamma^\mu\partial_\mu -
m(t)]\psi(x) = 0, \nonumber\\ \bar \psi(x)[i
\gamma^\mu\overleftarrow{\partial}_\mu + m(t)] = 0~,
\end{eqnarray}
where $\bar \psi = \psi^{\dagger}\gamma^0$. The corresponding
Hamiltonian is $(k = 1,2,3)$
\begin{equation} \label{Fermion_Hamiltonian} H(t) = i \int d^3x\ \psi^{\dagger}\dot
\psi = \int d^3x\ \bar \psi \{ -i \gamma^k\partial_k + m(t) \}
\psi.
\end{equation}
By analogy to the scalar case, we use the following decompositions
of field functions in the discrete momentum space:
\begin{eqnarray} \label{decomposition} \psi(x) =
\frac{1}{\sqrt{V}}\sum_{\mathbf{p}}\sum_{\alpha = 1,2}\bigl\{e^{i
\mathbf{p}\mathbf{x}}\, a_\alpha(\mathbf{p},
t)\,\mathsf{u}^{\alpha}(\mathbf{p}, t) + e^{-i
\mathbf{p}\mathbf{x}} \,b^{\dagger}_\alpha(\mathbf{p},
t)\,\mathsf{v}^{\alpha}(\mathbf{p}, t) \bigr\},\nonumber
\\[6pt]
\bar \psi(x) = \frac{1}{\sqrt{V}}\sum_{\mathbf{p}}\sum_{\alpha =
1,2}\bigl\{e^{-i \mathbf{p}\mathbf{x}}\,
a^{\dagger}_\alpha(\mathbf{p}, t)\,\bar{
\mathsf{u}}^{\alpha}(\mathbf{p}, t) + e^{i \mathbf{p}\mathbf{x}}
b_\alpha(\mathbf{p}, t)\,\bar{\mathsf{v}}^{\alpha}(\mathbf{p}, t)
\bigr\}.
\end{eqnarray}
The OR is intended to  derive  equations of motion for creation
and annihilation operators. It is based on the primary equations
(\ref{eq_of_mot_w_alt_mass}) and  free $u,\ v$ - spinors  with the
substitution $m \to m(t)$. Thus, the following  equations for the
spinors are postulated in the OR~:
\begin{eqnarray}
\left[\gamma p - m(t)\right] \mathsf{u}(\mathbf{p}, t) &=& 0,\nonumber \\
\left[\gamma p + m(t)\right] \mathsf{v}(\mathbf{p}, t) &=& 0
\end{eqnarray}
with $p^0 = \omega(\mathbf{p}, t)$. These definitions create the
set of standard orthogonality conditions \cite{BSh}  depending
 on time now parameterically
\begin{gather}
 \bar{\mathsf{u}}^{\alpha}(\mathbf{p}, t)\,
\mathsf{u}^{\beta}(\mathbf{p}, t) = \frac{m(t)}{\omega(\mathbf{p},
t)} \delta_{\alpha\beta}, \quad\quad
\bar{\mathsf{v}}^{\alpha}(\mathbf{p}, t)
\,\mathsf{v}^{\beta}(\mathbf{p}, t) =
-\frac{m(t)}{\omega(\mathbf{p}, t)}\,\delta_{\alpha\beta},
\nonumber \\ \mathsf{u}^{\dagger \alpha}(\mathbf{p}, t)
\,\mathsf{u}^{\beta}(\mathbf{p}, t)= \mathsf{v}^{\dagger
\alpha}(-\mathbf{p},t)
\,\mathsf{v}^{\beta}(-\mathbf{p}, t) = \delta_{\alpha\beta},  \nonumber \\
\bar{\mathsf{u}}^{\alpha}(\mathbf{p}, t)\,
\mathsf{v}^{\beta}(\mathbf{p}, t) = {\mathsf{u}}^{\dagger
\alpha}(\mathbf{p}, t)\, \mathsf{v}^{\beta} (-\mathbf{p}, t) =0
.\label{cond_of_ortonorm}
\end{gather}

Decompositions (\ref{decomposition}) and  relations
(\ref{cond_of_ortonorm}) lead immediately  to the diagonal  form
of the Hamiltonian (\ref{Fermion_Hamiltonian})
\begin{equation}
\label{diagonal_Ham} H(t) = \sum_{\mathbf{p},\alpha} \omega
(\mathbf{p},t)[a^{\dagger}_{\alpha} (\mathbf{p},t)
a_{\alpha}(\mathbf{p},t) + b^{\dagger}_{\alpha} (\mathbf{p},t)
b_{\alpha} (\mathbf{p},t)]
\end{equation}
with interpretation of $a^{\dagger},\ a$ (and $b^{\dagger},\ b$)
as the creation and annihilation operators of quasiparticles
obeying the standard anti-commutation relations
\begin{equation}\label{10}
\{a_{\alpha}(\mathbf{p},t), a^{\dagger}_{\beta}(\mathbf{p}', t)\}
= \{b_{\alpha}(\mathbf{p},t), b^{\dagger}_{\beta}(\mathbf{p}',
t)\} = \delta_{\mathbf{p}\mathbf{p}'}\delta_{\alpha\beta}.
\end{equation}
We are not interested in the subsequent diagonalization of the
spin operator and such QPR can be named the {\em incomplete}
representation.

Now in order to get equations of motion for creation and
annihilation operators  in the OR, let us substitute the
decomposition (\ref{decomposition}) in Eqs.
(\ref{eq_of_mot_w_alt_mass}) and use relations
(\ref{cond_of_ortonorm}). Then, as an intermediate result, we
obtain the following closed set of equations of motion which is
valid in a general case
\begin{align}
\label{matrix_eq_of_mot} \dot{a}_{\alpha}(\mathbf{p}, t) &+
U_{1}^{\alpha\beta}(\mathbf{p}, t)\, a_{\beta}(\mathbf{p}, t) +
U_{2}^{\alpha\beta}(\mathbf{p}, t)\,
b^{\dagger}_{\beta}(-\mathbf{p}, t) = -i \omega(\mathbf{p}, t)\,
a_{\alpha}(\mathbf{p}, t),\nonumber
\\[5pt]
\dot{a}^{\dagger}_{\alpha}(\mathbf{p}, t) &-
a^{\dagger}_{\beta}(\mathbf{p}, t)\,
U_{1}^{\beta\alpha}(\mathbf{p}, t) + b_{\beta}(-\mathbf{p}, t)\,
U_{2}^{\beta\alpha}(\mathbf{p}, t) = i \omega(\mathbf{p}, t) \,
a^{\dagger}_{\alpha}(\mathbf{p}, t),\nonumber
\\[5pt]
\dot{b}_{\alpha}(-\mathbf{p}, t) & +
a^{\dagger}_{\beta}(\mathbf{p}, t)\,
V_{1}^{\beta\alpha}(\mathbf{p}, t) - b_{\beta}(-\mathbf{p}, t)\,
V_{2}^{\beta\alpha}(\mathbf{p}, t) = - i \omega(\mathbf{p}, t)\,
b_{\alpha}(-\mathbf{p}, t),\qquad \nonumber
\\[5pt]
\dot{b}^{\dagger}_{\alpha}(-\mathbf{p}, t) & +
V_{1}^{\alpha\beta}(\mathbf{p}, t)\, a_{\beta}(\mathbf{p}, t) +
V_{2}^{\alpha\beta}(\mathbf{p}, t) \,b_{\beta}(-\mathbf{p}, t) = i
\omega(\mathbf{p}, t) b^{\dagger}_{\alpha}(-\mathbf{p},t).
\end{align}
The spinor construction is introduced here as
\begin{align}
\label{spinor_constr} U^{\alpha\beta}_{1} &=
\mathsf{u}^{\dagger\alpha}(\mathbf{p}, t)\,
\dot{\mathsf{u}}^{\beta}(\mathbf{p}, t), & V^{\alpha\beta}_{1} &=
\mathsf{v}^{\dagger\alpha}(-\mathbf{p}, t)
\,\dot{\mathsf{u}}^{\beta}(\mathbf{p}, t), \nonumber \\[5pt]
U^{\alpha\beta}_{2} &= \mathsf{u}^{\dagger\alpha}(\mathbf{p},
t)\,\dot{ \mathsf{v}}^{\beta}(-\mathbf{p}, t), &
V^{\alpha\beta}_{2} &= \mathsf{v}^{\dagger\alpha}(-\mathbf{p},
t)\,\dot{\mathsf{v}}^{\beta}(-\mathbf{p}, t).
\end{align}
The matrices $U_{2}$ and $V_{1}$ describe the transitions between
states with  positive and negative energies and different spins,
while the antiunitary matrices $U_{1}$ and $V_{2}$ show the spin
rotations only
\begin{equation}\label{matr}
    U^{\dagger}_{1} = - U_{1}, \qquad V^{\dagger}_{2}  = -
    V_{2}, \qquad V^{\dagger}_{2}  = - U_{2}.
\end{equation}
Equations (\ref{matrix_eq_of_mot}) are compatible with the
canonical commutation relations (\ref{10}).

Let us write now the $u,\ v$ - spinors in an explicit form,
according to \cite{Rayder}:
\begin{align} \label{free_spinors}
\mathsf{u}^{\dagger \,1}(\mathbf{p}, t) &= A(\textbf{p})\, [\,\omega_+, 0, p_3,p_-],\nonumber\\
\mathsf{u}^{\dagger \,2}(\mathbf{p}, t) &= A(\mathbf{p})\,[\,0,\omega_+, p_+, -p_3\,], \nonumber\\
\mathsf{v}^{\dagger \,1}(-\mathbf{p}, t) &= A(\textbf{p})\,[-p_3, -p_-, \omega_+, 0\,], \nonumber\\
\mathsf{v}^{\dagger \,2}(-\mathbf{p}, t) &= A(\textbf{p})\, [-p_+,
p_3, 0, \omega_+],
\end{align}
where $p_\pm = p_1 \pm i p_2,\ \omega_+ = \omega + m(t)$ and
$A(\textbf{p}) = [2\omega\omega_+]^{-1/2}$. Spin rotation matrices
(\ref{spinor_constr}) in this representation are equal to zero
\begin{equation}
\label{spin_rot_matr} U_{1} = V_{2} = 0.
\end{equation}
For the remaining matrices (\ref{spinor_constr}) we have $U_{2} =
- V_{1} = U$, where $U$ is the hermitian matrix
\begin{equation}\label{45}
U(\mathbf{p}, t) =   \frac{\dot m(t)}{2\omega^2(\mathbf{p}, t)}
\begin{bmatrix}
    \  p_3 & p_- \\
    \  p_+ & -p_3 \\
    \end{bmatrix}
\end{equation}
Thus, the system of equations of motion (\ref{matrix_eq_of_mot})
reduces to the following one:
\begin{eqnarray}
\dot a_{\alpha}(\mathbf{p}, t) + U_{\alpha\beta}(\mathbf{p}, t)
b^{\dagger}_{\beta}(-\mathbf{p}, t) &=& -i\omega(\mathbf{p},
t)a_{\alpha}(\mathbf{p}, t),
\nonumber \\
\dot b_{\alpha}(-\mathbf{p}, t) - a^{\dagger}_{\beta}(\mathbf{p},
t) U_{\beta\alpha}(\mathbf{p}, t) &=& -i\omega(\mathbf{p},
t)b_{\alpha}(-\mathbf{p}, t). \label{red_matrix_eq_of_mot}
\end{eqnarray}

\subsection{Kinetic equation \label{secf2}}

Equations of motion (\ref{red_matrix_eq_of_mot}) do not contain
the spin rotation matrices (\ref{spin_rot_matr}) and they are
similar to Eqs. (\ref{eq_of_mot}); therefore, the KE derivation
meets no problem now. To be specific, let us introduce the
one-particle correlation functions
\begin{eqnarray}\label{corr}
g_{\alpha\beta}(\mathbf{p}, t) &=& \langle
0|a_{\beta}^{\dagger}(\mathbf{p}, t)a_{\alpha}(\mathbf{p},
t)|0 \rangle,\nonumber \\
\tilde{g}_{\alpha\beta}(\mathbf{p}, t) &=& \langle
0|b_\beta(-\mathbf{p}, t)b^{\dagger}_\alpha(-\mathbf{p},
t)|0\rangle.
\end{eqnarray}
The differentiation of (\ref{corr}) with respect to time leads to
 the following matrix equations
\begin{eqnarray}
\label{corr_58} \dot{g}(\mathbf{p}, t) &=& -U(\mathbf{p}, t)
\,G(\mathbf{p}, t) -
G^{\dagger}(\mathbf{p}, t)U(\mathbf{p}, t),\nonumber \\
\dot{\tilde{g}}(\mathbf{p}, t) &=& G(\mathbf{p}, t) U(\mathbf{p},
t) + U(\mathbf{p}, t)\, G^{\dagger}(\mathbf{p}, t)~,
\label{corr_eq_of_mot}
\end{eqnarray}
 where the auxiliary function was introduced
\begin{eqnarray}
\label{aux_corr} G_{\alpha\beta}(\mathbf{p}, t) = \langle 0|
a_\beta^{\dagger}(\mathbf{p}, t) b_\alpha^{\dagger}(-\mathbf{p},
t) |0 \rangle .
\end{eqnarray}
Together  with Eqs. (\ref{corr_58}), the corresponding equation of
motion
\begin{eqnarray}
\dot{G}(\mathbf{p}, t) &=& U(\mathbf{p}, t)g(\mathbf{p}, t) -
\tilde g(\mathbf{p}, t) U(\mathbf{p}, t) + 2i\omega(\mathbf{p}, t)
G(\mathbf{p}, t) \label{aux_eq_of_mot}
\end{eqnarray}
 forms a closed set of equations
for correlation functions. With  the help of Eqs.
(\ref{aux_eq_of_mot}), one can exclude the auxiliary correlator
from the system (\ref{corr_eq_of_mot})
\begin{align}
\dot{g}(\mathbf{p}, t) &= 2 U(\mathbf{p}, t) \int\limits_{t_0}^t
dt' [ \,\tilde{g} (\mathbf{p}, t') U(\mathbf{p}, t') -
U(\mathbf{p}, t')g(\mathbf{p}, t')]
\cos{{2\theta(\mathbf{p},t',t)}},\nonumber
\\
\dot{\tilde{g}}(\mathbf{p}, t) &= 2\int\limits_{t_0}^t dt' [
U(\mathbf{p}, t')g(\mathbf{p}, t') - \,\tilde{g} (\mathbf{p}, t')
U(\mathbf{p}, t')] U(\mathbf{p}, t)
\cos{{2\theta(\mathbf{p},t',t)}}, \label{g_eq_of_mot}
\end{align}
using zero initial conditions. The subsequent transformation is
based on the relation
\begin{equation}
Tr \{U(t)A U(t')\} = \frac{1}{4} \lambda(\mathbf{p},
t)\lambda(\mathbf{p}, t') Tr A, \label{60}
\end{equation}
for an arbitrary second-rank matrix $A$, which follows from Eq.
(\ref{45}). The function $(p=|\mathbf{p}|)$
\begin{equation}
\lambda(\mathbf{p}, t) = \frac{\dot{m}(t)p}{\omega^2(\mathbf{p},
t)} \label{lmbd}
\end{equation}
plays a role of some analog of Eq. (\ref{Delta}).

Using  isotropy of the considered system,  we will limited
ourselves to the spin-averaged scalar distributions
\begin{equation}
f(\mathbf{p}, t) = \frac12 Tr\ g(\mathbf{p}, t),\qquad
\tilde{f}(-\mathbf{p}, t) = 1- \frac12 Tr\ \tilde{g}(\mathbf{p},
t) .
\end{equation}
Calculating the trace of (\ref{g_eq_of_mot}) we get
\begin{multline}
\dot{f}(\mathbf{p}, t) = \dot{\tilde{f}}(-\mathbf{p}, t) = \frac12
\lambda(\mathbf{p}, t) \int\limits_{t_0}^{t} dt'
\lambda(\mathbf{p}, t') [1- \tilde{f}(-\mathbf{p}, t') \\ -
{f}(\mathbf{p}, t')] \cos{2\theta(\mathbf{p}; t,t')}.
\end{multline}
In  the case of the vacuum initial state, $\tilde{f}(-\mathbf{p},
t)={f}(\mathbf{p}, t)$, we have
\begin{equation} \label{FermKE} \dot{f}(\mathbf{p},
t) = 2\lambda(\mathbf{p}, t) \int\limits_{t_0}^t dt'
\lambda(\mathbf{p}, t')[1 - 2f(\mathbf{p}, t')] \cos[2\theta(t,
t')],
\end{equation}
The KE's (\ref{I_KE}) and (\ref{FermKE}) are similar but differ by
statistical factors $1 \pm 2f$ (the Bose enhancement or the Fermi
suppression) and by structure of the factors (\ref{Delta}) and
(\ref{lmbd}). The corresponding linear equations for the
non-Hamiltonian dynamical system become
\begin{equation}\label{odu} \dot f = \frac{1}{2}\lambda
u,\qquad \dot u = \lambda[1 - 2f] - 2\omega v,\qquad \dot v =
2\omega u.
\end{equation}
This system possesses one first integral of motion (see
\cite{Vinnik})
\begin{equation}\label{f13}
 (1-2f)^{2}+v^{2}+u^{2}= 1.
\end{equation}
This relation represents an ellipsoid in the phase space of
$(f,u,v)$ variables. After exclusion of the function $f$ from Eq.
(\ref{odu}), we obtain the system of non-linear equations
\begin{eqnarray}
  \dot{u} &=& \lambda\sqrt{1-u^{2}-v^{2}}-2\omega v , \nonumber \\
  \dot{v} &=& 2\omega u.
\label{f16}
\end{eqnarray}
It can easily be proved that the KE (\ref{FermKE}) is invariant
with respect to time inversion. Equations analogous to Eqs.
(\ref{ODE_SYS}) and (\ref{odu}) were obtained in \cite{49} (see
also \cite{Grib}) for the conformal flat space-time.  The KE
(\ref{FermKE}) to the case of spinor QED was generalized in
\cite{SPIE05,Skokov}. In the general case, the spin correlation
functions (\ref{corr}) and (\ref{aux_corr})  can be decomposed in
respect of the Pauli matrices ({\em e.g.}, \cite{Silin}).

\subsection{Observables and regularization \label{secf3}}

The total particle number density and energy density in the
considered case are distinguished from the corresponding
expressions (\ref{dens}) and (\ref{t00}) for the scalar system by
the spin degeneration factor $g=4$ (for an equal number of
particles and antiparticles)
\begin{eqnarray}
\label{f_density}
n(t) = 4 \int [dp] f(\mathbf{p}, t),\\
\epsilon(t) = \langle0|T_{00}|0\rangle = 4 \int [dp]
\omega(\mathbf{p}, t) f(\mathbf{p}, t)~, \label{f_energy}
\end{eqnarray}
where $T_{00}$ is the zero component of the energy-momentum tensor
\begin{equation}
T_{\mu\nu} = \frac{i}{2} [\bar{\psi} \gamma_{\mu} (\partial_{\nu}\psi) -
(\partial_{\nu}\bar{\psi})\gamma_{\mu}\psi].
\label{f_Tensor}
\end{equation}
By definition, the entropy density of the fermion system is equal
to
\begin{equation}
S(t) = - 4 \int [dp] \bigl[f\ln{f} + (1-f)\ln{(1-f)}
\bigr].\label{f_entropy}
\end{equation}
Finally, the pressure is
\begin{equation}
P (t) = \frac{1}{3} <T_{kk}>. \label{f_Press_gen}
\end{equation}
Using (\ref{f_Tensor}) and (\ref{eq_of_mot_w_alt_mass}), this
relation can be reduced to the following form:
\begin{equation}
P (t) = \frac{1}{3}\{\epsilon(t) - m(t)
\langle\bar{\psi}(x)\psi(x)\rangle \}.
\end{equation}
Here the correlation function is calculated by means of relations
(\ref{cond_of_ortonorm})
\begin{equation}
P (t) = \frac{1}{3}\epsilon(t) + \frac{1}{3} \int [dp]\
\frac{m^2(t)}{\omega(\mathbf{p}, t)}\ [1- 2f(\mathbf{p}, t)] +
P_{pol}(t)~,
\end{equation}
where the last term takes into account the contribution of the
vacuum polarization,
\begin{equation}
P_{pol}(t) = -\frac{2}{3} m(t) \int [dp]\
\frac{p}{\omega(\mathbf{p}, t)}\ u(\mathbf{p}, t).
\end{equation}
This result was obtained under additional conditions for
"observable" correlation functions
\begin{eqnarray}
\langle 0|a^{\dagger}_{\alpha}(\mathbf{p}, t)
a_{\beta}(\mathbf{p}', t) |0 \rangle = \langle
0|b^{\dagger}_{\alpha}(\mathbf{p}, t) b_{\beta}(\mathbf{p}',
 t) |0 \rangle =
 \delta_{\alpha\beta}\delta_{\mathbf{p}\mathbf{p}'}~,
\end{eqnarray}
which is a consequence of space-homogeneity  and isotropy (absence
of the spin moment)  of the system. The auxiliary correlation
function (\ref{aux_corr}) is not connected with the spin moment
and, therefore, it remains off-diagonal with respect to spin
indices.

The same regularization procedure can be realized here for the
calculation of divergence integrals, as presented in Sect.
\ref{SF3} for the scalar bosons. Equation (\ref{star}) for the
distribution function in the low-density approximation is valid
also after the replacement $\Delta (\mathbf{p},t) \to \lambda
(\mathbf{p},t)$. Asymptotics of the factor (\ref{lmbd}) is equal
to $\lambda^{\infty} (\mathbf{p},t) = \dot{m} (t)/p$. Using Eq.
(\ref{star}) and the rules of Sect. \ref{SF3}, one can derive the
following expressions for counterterms for the case of fermion
fields:
\begin{align}
f_M^{(4)} (\mathbf{p},t) &= \left[\frac{\dot{m}(t)}{4(p^2 + M^2)}\right]^2, \nonumber\\
u_M^{(3)} (\mathbf{p},t) &= \frac{\ddot{m}(t)}{4(p^2 + M^2)^{3/2}}.
\end{align}
However, these counterterms can be ignored in computer
calculations.

It is known \cite{Grib} that the phase density of pairs created in
an electric field for the whole period of its action is related to
long-time asymptotics of solutions of some oscillator equations.
For the inertial mechanism,  analogous derivation results in the
following relation \cite{Andreev02}
\begin{align}\label{A23}
\lim\limits_{t\to +\infty} f(\mathbf{p},t) = \frac{\cosh{\tau_{\pi}[(m_i -
m_f)]}-\cosh{\tau_{\pi}[ (\omega_i - \omega_f)}]
}{2\sinh{(\tau_{\pi}\omega_i)}\sinh{(\tau_{\pi}\omega_f)}}, 
\end{align}
where $\tau_{\pi}= \pi \tau /2$, the indices $i,f$ correspond to
the initial and final states. This relation is convenient for
calculation of observables for long pulses $\tau \gg m$, {\em
e.g.} Fig. \ref{Energy} where the direct solution of kinetic
equation is a very robust numerical problem.

\subsection{Numerical results \label{secf4}}
Here  numerical investigations of the KE's are presented for
bosons (\ref{I_KE}), fermions (\ref{FermKE}) and appropriate
densities of observable variables  are estimated (Sects. \ref{SF3}
and \ref{secf3}).  As an example, two variants of time-dependent
masses are considered. The first case qualitatively corresponds to
a typical meson mass change under the phase transition within the
NJL model \cite{Rehberg}
\begin{equation}\label{test_mass_law1}
m(t) = (m_0 - m_f) \exp{[-(t/\tau)^2]} + m_f, \qquad t \ge 0,
\end{equation}
with the parameters $m_0$ (initial mass), $m_f$ (final mass), and
$\tau$ (transition time). Another variant suggested in
\cite{Andreev02,Andreev96} allows  an analytical solution of the
Dirac equation
\begin{equation} \label{test_mass_law2}
 m(t) = \frac{m_f + m_0}{2} + \left(\frac{m_f -
 m_0}{2}\right) \tanh{(2t/\tau)}.
\end{equation}

\begin{figure}[t]
\centering
\includegraphics[width=0.49\textwidth,keepaspectratio]{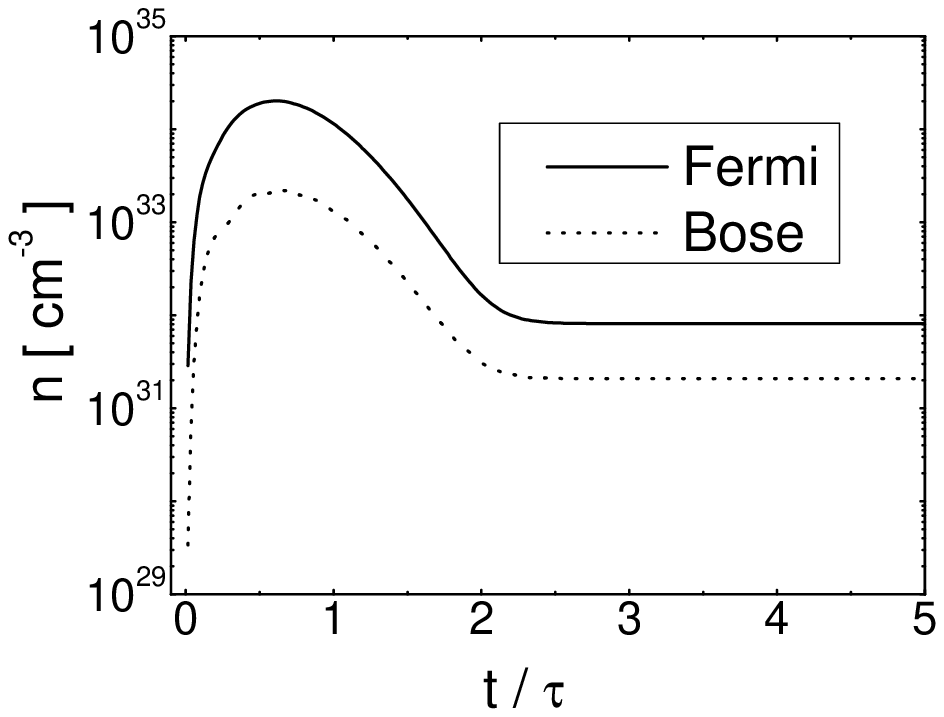}
\includegraphics[width=0.49\textwidth,keepaspectratio]{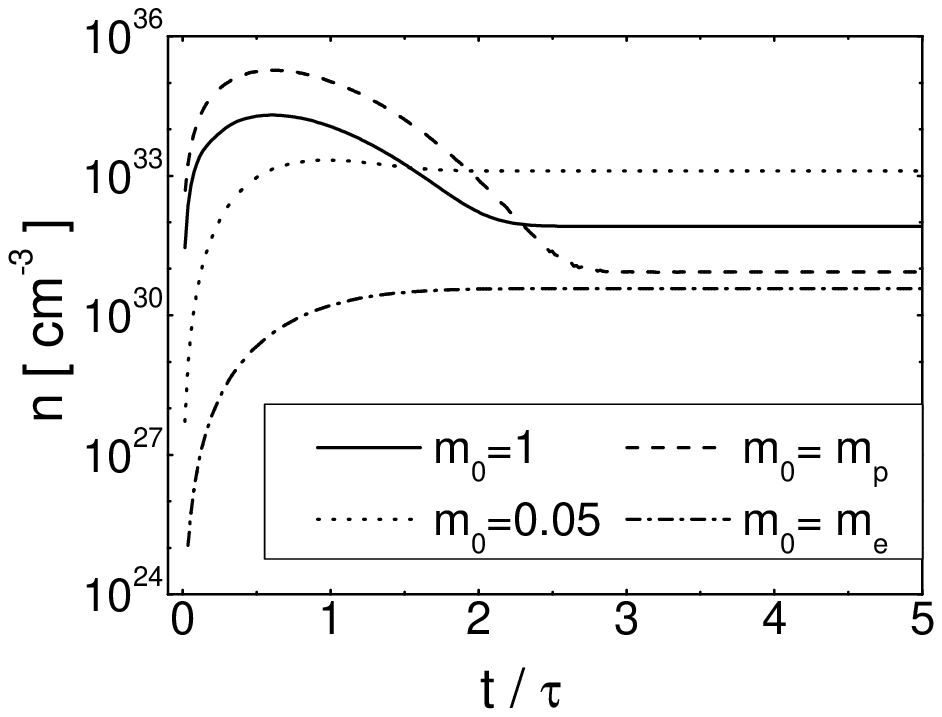}
\parbox[t]{0.48\textwidth}{
\caption{Time dependence of the pair density  for bosons and
fermions at $m_0=1$.} \label{fig_2}} \hspace{2.5mm}
\parbox[t]{0.48\textwidth}{
\caption{The pair density evolution of fermions for various
initial
masses\newline (p - proton, e - electron).} \label{fig_3}}\\[15pt]
%
\includegraphics[width=0.49\textwidth,keepaspectratio]{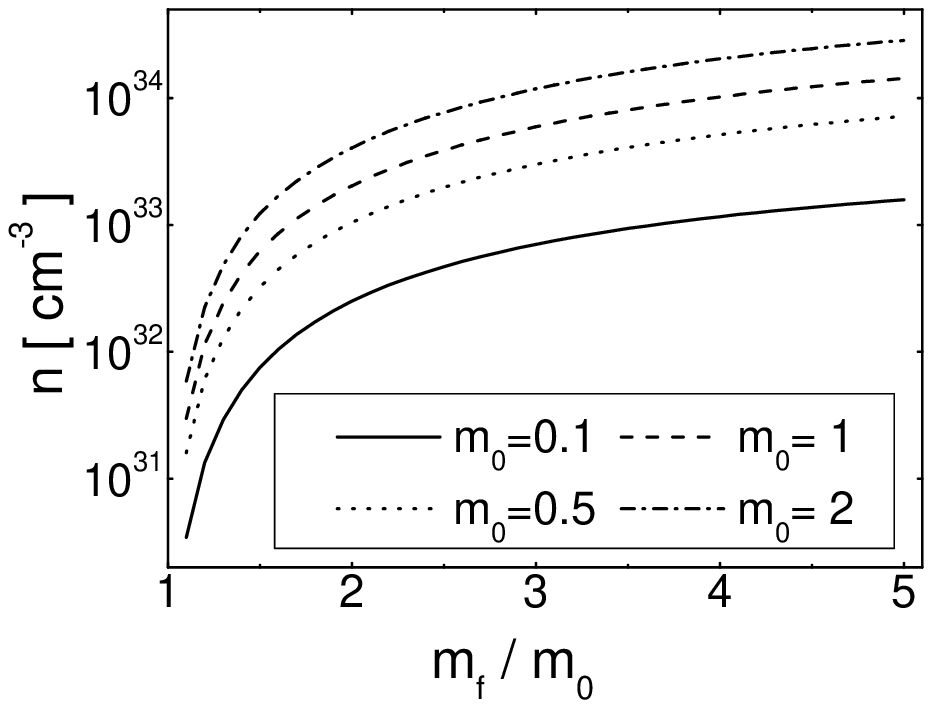}
\includegraphics[width=0.49\textwidth,keepaspectratio]{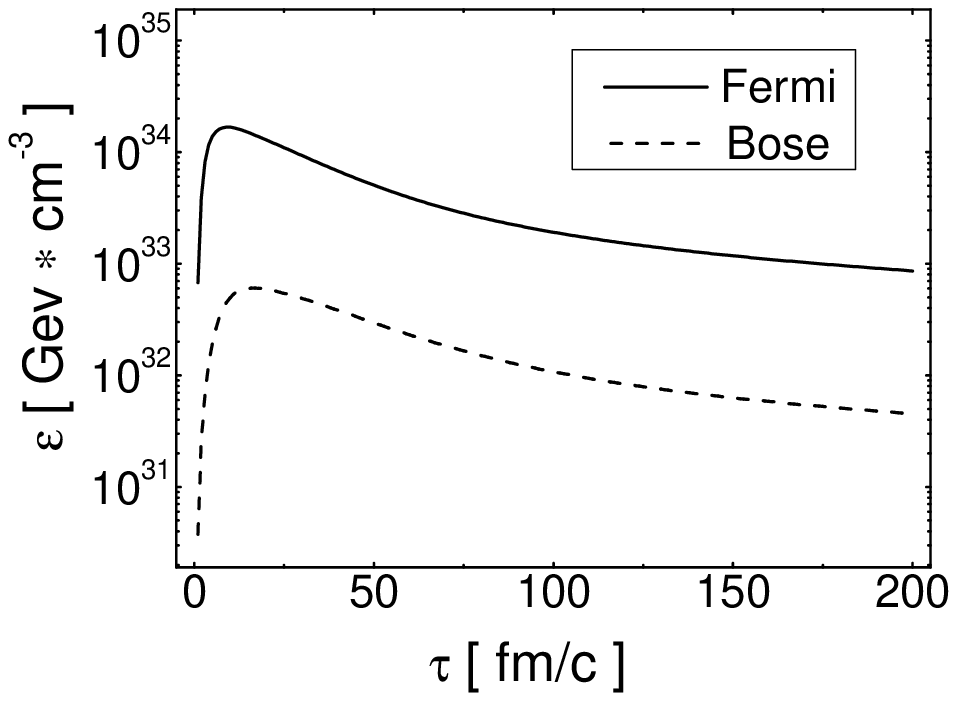}
\parbox[t]{0.48\textwidth}{
\caption{The residual pair density  versus the mass ratio.}
\label{fig_2.2.7}} \hspace{2.5mm}
\parbox[t]{0.48\textwidth}{
\caption{The residual energy density versus the transition time
$\tau$, $m_0=1$.}\label{Energy}}
\end{figure}

\begin{figure}[t]
\centering
\includegraphics[width=0.48\textwidth,keepaspectratio]{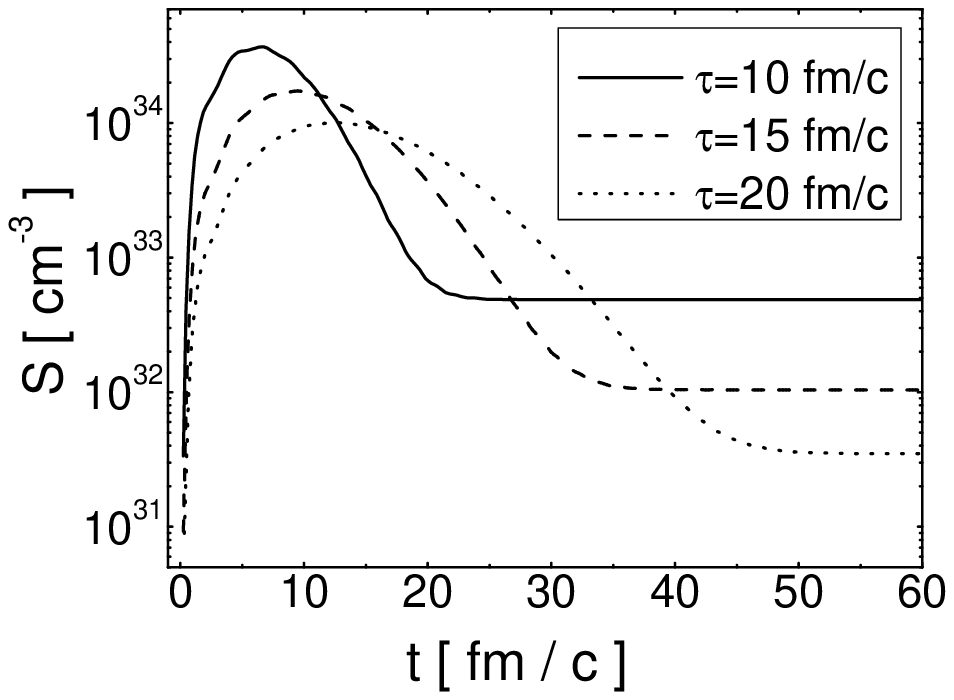}
\includegraphics[width=0.48\textwidth,height=45mm]{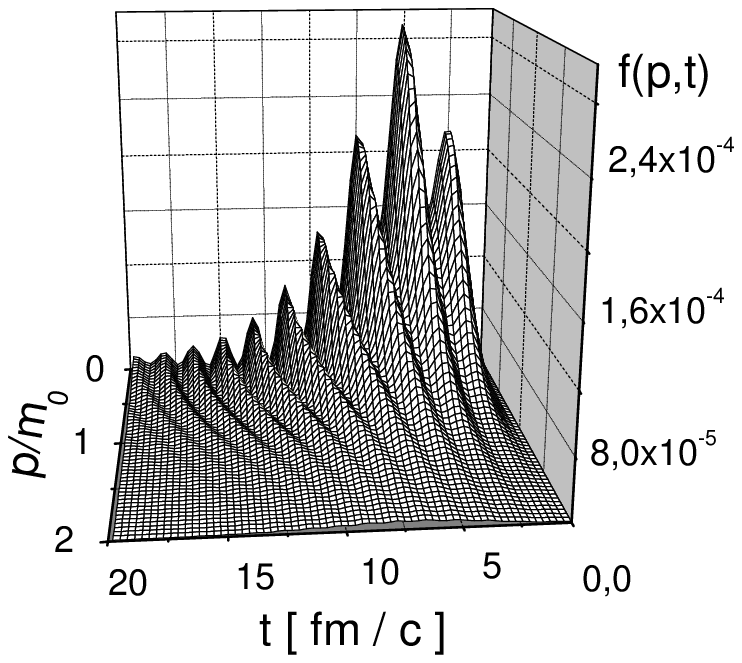}
\parbox[t]{0.48\textwidth}{
\caption{Entropy density evolution of bosons for  different values
of the relaxation time $\tau$, $m_0=1$.} \label{fig_4}}
\hspace{2.5mm}
\parbox[t]{0.48\textwidth}{
\caption{The distribution function of bosons in time-momentum
variables, $m_0=1$, $\tau=10$~fm/c.}\label{fig_3.2.6}}\\[15pt]
%
\includegraphics[width=0.48\textwidth,keepaspectratio]{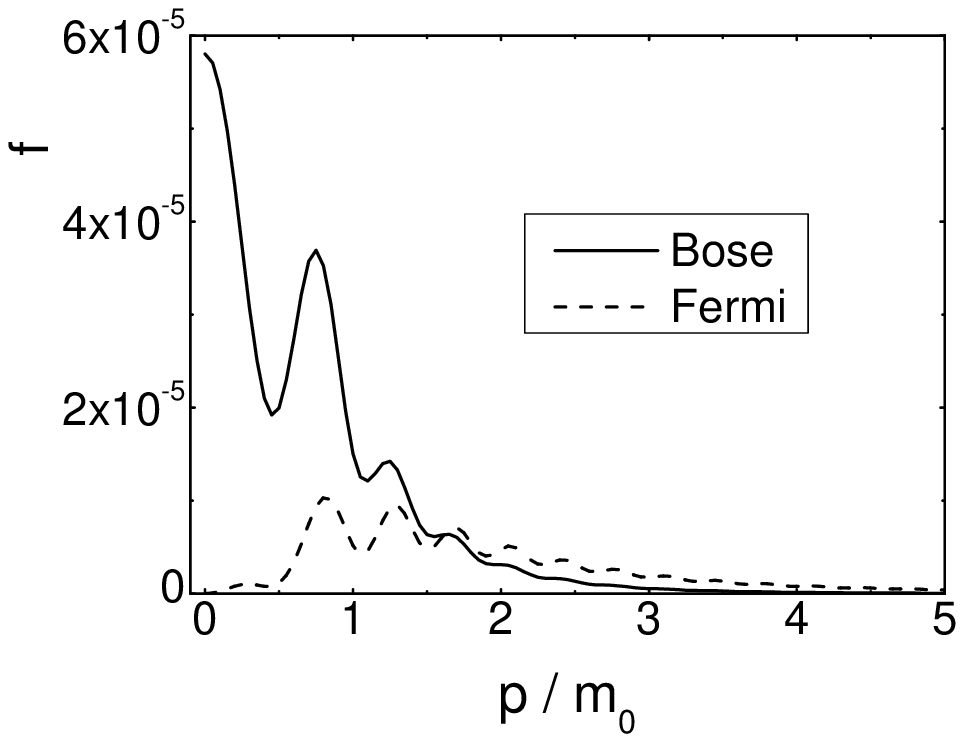}
\includegraphics[width=0.48\textwidth,keepaspectratio]{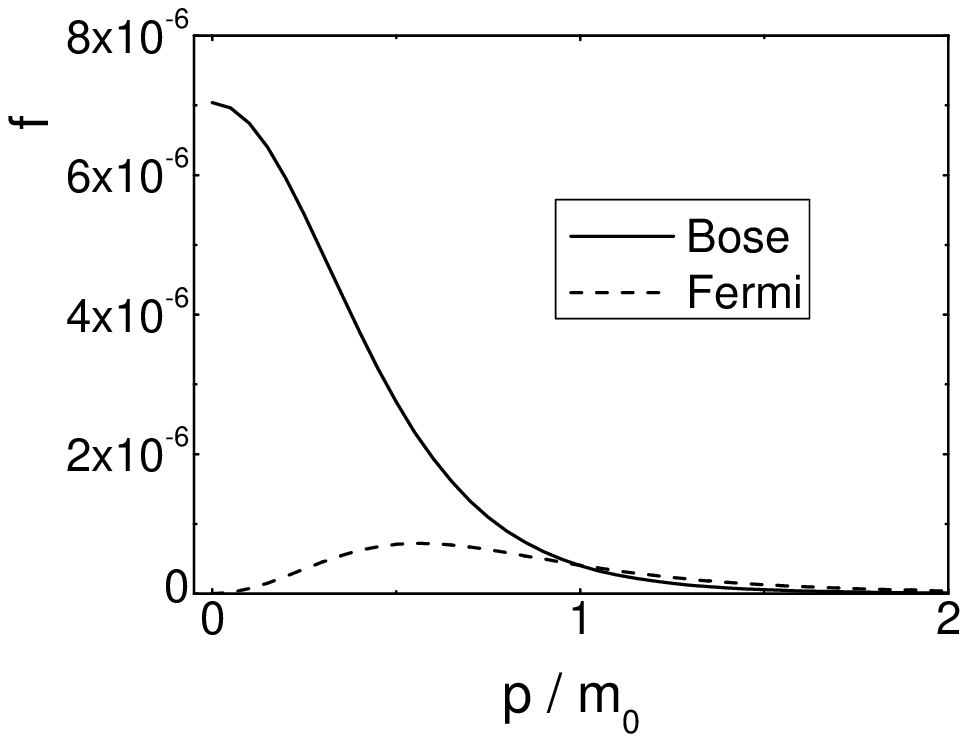}
\parbox[t]{0.48\textwidth}{
\caption{Distribution functions at the time  $t = \tau=10$~fm/c,
$m_0=1$.} \label{fig_5}} \hspace{2.5mm}
\parbox[t]{0.48\textwidth}{
\caption{Asymptotic form of distribution functions at $t \gg
\tau$.} \label{fig_6}}
\end{figure}

\begin{figure}[h]
\centering
\includegraphics[width=0.48\textwidth,keepaspectratio]{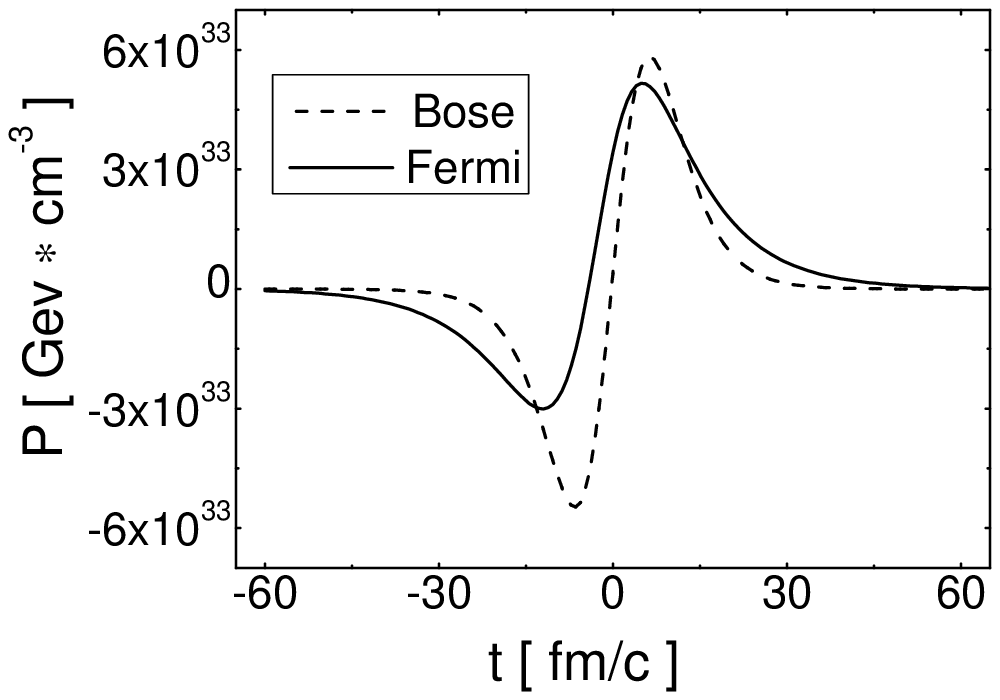}
\includegraphics[width=0.48\textwidth,keepaspectratio]{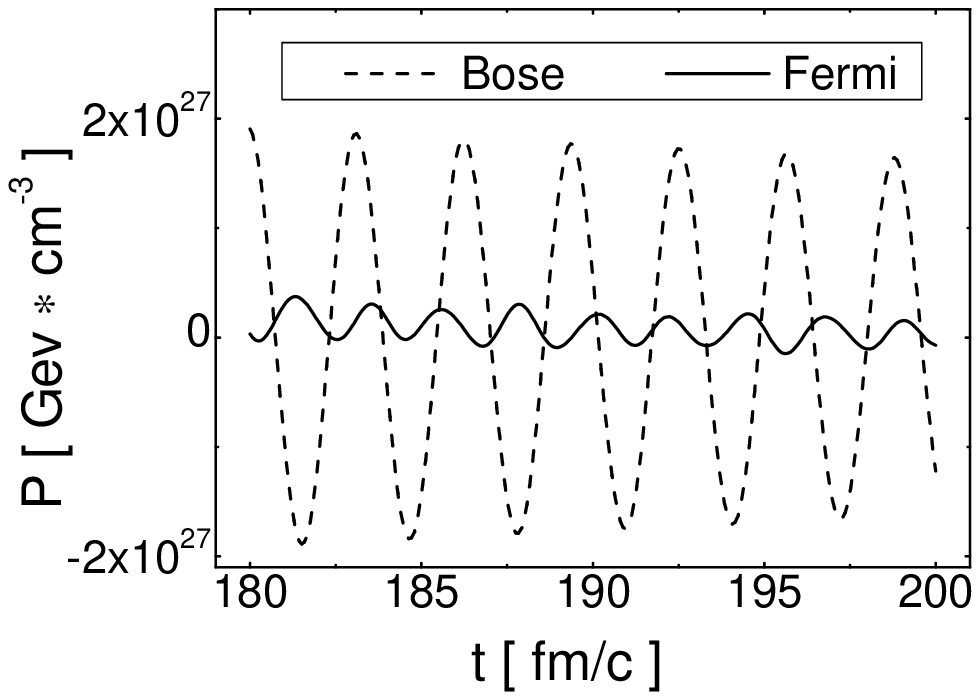}
\parbox[t]{0.48\textwidth}{
\caption{Pressure evolution corresponding to mass changing
(\ref{test_mass_law2}), $m_0=1$, $\tau = 20$~fm/c.} \label{fig_7}}
\hspace{2.5mm}
\parbox[t]{0.48\textwidth}{
\caption{Asymptotic oscillations of pressure at $t \gg \tau$
without `switching off' of vacuum polarization effects.
\label{fig_8}}}
\end{figure}

Numerical results for solution of the KEs are presented in Figs.
\ref{fig_2}-\ref{fig_6} for $m(t)$ defined by Eq.
(\ref{test_mass_law1}) with $\tau=10 \ fm/c$ and in Figs.
\ref{fig_7},\ref{fig_8} for the mass (\ref{test_mass_law2}) with
$\tau = 20 \ fm/c$.  Masses are specified in natural units,
$m_h=197$ MeV$/c^2$.

At a glimpse the time dependence of quasiparticle density for
bosons and fermions repeats qualitatively the curve $\dot{m}(t)$;
however, when $\dot{m}(t)\to 0$ the densities go asymptotically to
certain finite values $n_r$ (residual density, Fig. \ref{fig_2},
$t\gg\tau$), which characterize the real (free) particles (at the
active stage of the process, $\dot{m}(t) \neq 0$, one may talk
about quasiparticles only). The  $n(t)$ dependence for fermions on
the initial mass value  is shown in Fig. \ref{fig_3} in the range
from the electron mass to proton one. This dependence is
nonmonotonic: With increasing $m_0$  the residual density reaches
the maximum at $m_0 \sim 10$ MeV and then begins to decrease. This
effect appreciably depends on the variant used for the mass
change: For the case (\ref{test_mass_law2}) it manifests itself
much more clearly than for the model (\ref{test_mass_law1}). The
residual particle density dependence on the mass change is
presented in Fig. \ref{fig_2.2.7}.
Qualitative behavior of the energy density is similar to that of
particle density, showing a smooth asymptotic decrease (see Fig.
\ref{Energy}). In Fig. \ref{fig_4}  the time dependence of the
boson entropy density  is presented for different values of the
relaxation time. Non-monotonic  behavior of entropy is caused by
the fact that the system is opened and treated in the
non-dissipative approximation.

 Momentum spectra of particles at different stages of the
interaction process are shown in Figs.
\ref{fig_3.2.6}-\ref{fig_6}. The maximal number of bosons is
created with zero momentum, whereas there are no fermions with
$p=0$. This feature differs qualitatively from the case of the
Schwinger mechanism of particle creation \cite{Vinnik}.

It is important that the formation of appreciably non-monotonic
distributions with the "fast" mass  changing
(\ref{test_mass_law1}), Fig. \ref{fig_5}, assists in the
development of plasma oscillations. For smoother mass changing
(\ref{test_mass_law2}) this effect becomes much less pronounced.

The most interesting features are observed in the pressure
behavior, Fig. \ref{fig_7}-\ref{fig_8}.  For both variants of the
mass evolution and independently of particle statistics, the
pressure is negative at the beginning of process, then it changes
its sign in the reflection point of $m(t)$ and  gradually
decreases. However, contrary to other observables, the pressure
has no constant asymptotics and at $t \gg\ \tau$  looks like
almost un-damped oscillations, Fig. \ref{fig_7}. It is distinctive
for pressure of the bosonic quasiparticle system which strongly
oscillates around zero, Fig. \ref{fig_8}. The reason is that
unlike the other considered quantities, the pressure is not
completely determined by the quasiparticles distribution function
$f(\mathbf{p},t)$ , but it depends also on the function
$u(\mathbf{p},t)$, which describes vacuum polarization effects. At
the operator language, this means incomplete diagonalization of
the energy-momentum tensor in the Fock space:  Averaged over the
initial vacuum, its spatial components include the contribution of
anomalous correlators like $\langle 0\, | a_p^{\dagger}\,
a_p^{\dagger} \,| 0 \rangle$.

Thus, if the process of particle creation stops when the time mass
evolution  is completed ($\dot{m}(t)\to 0$), the vacuum
polarization effects are not "switched off" simultaneously but
continue to influence some observables, {\em e.g.},  pressure. As
a consequence, in such non-dissipative nonequilibrium model it is
impossible to determine unambiguously the equation of state
\cite{DM2006}.
\section{Massive vector bosons \label{MVB}}

\subsection{The complete QPR \label{MVB1}}

The simplest version of quantum field theory of neutral massive
vector bosons is given by the Lagrangian density
\cite{BSh}
\begin{equation} \label{Lagr_dens} \mathcal{L}(x) = - \frac{1}{2}\, \partial_\mu
u_{\nu}\, \partial^\mu u^\nu + \frac{1}{2}\,  m^2(t) u_\nu u^\nu ,
\end{equation}
which corresponds to the  equation of motion
\begin{equation}
\label{eq2a} [ \partial_{\mu}\partial^{\mu} + m^2(t) ] u_\nu = 0
\end{equation}
with the additional "external" constraint
\begin{equation}\label{add}
\partial_\mu u^\mu=0.
\end{equation}
An alternative way is to proceed from the Wentzel Lagrangian
\cite{Went}
\begin{equation}\label{lagr_w}
  \mathcal{L}(x) = -\frac14 F^{\mu\nu}F_{\mu\nu} + \frac12 m^2(t) u_\nu
  u^\nu,
\end{equation}
where the strength tensor $F_{\mu\nu}=\partial_\nu u_\mu -
\partial_\mu u_\nu$. Here the constraint (\ref{add}) is  a
consequence of dynamical equations. Lagrangians (\ref{Lagr_dens})
and (\ref{lagr_w}) result in different energy-momentum tensors
({\em e.g.}, see \cite{Went,Pavel}).

The transition to the QPR  is carried out by the standard
decomposition of free fields and momenta in the discrete momentum
space with the replacement $m\to m(t)$ in the dispersion law (see
Sect. \ref{2.1}),
\begin{align} u_\mu(x) &=
\frac{1}{\sqrt{2V}}\sum_{\mathbf{p}}\frac{1}{\sqrt{\omega(\mathbf{p},
t)}} \biggl\{a_\mu(\mathbf{p}, t)\, e^{i\mathbf{p} \mathbf{x}}  +
a^{*}_\mu(\mathbf{p}, t)\, e^{-i\mathbf{p} \mathbf{x}} \biggr\} , \nonumber \\
\pi_\mu(x) &= -
\frac{i}{\sqrt{2V}}\sum_{\mathbf{p}}\sqrt{{\omega(\mathbf{p}, t)}}
\left\{a_\mu(\mathbf{p}, t)\, e^{i\mathbf{p} \mathbf{x}} -
 a^{*}_\mu(\mathbf{p}, t)\, e^{-i\mathbf{p} \mathbf{x}} \right\},\label{or}
\end{align}
where $a_\mu(\mathbf{p}, t)$ are the classical amplitudes. Unlike
the scalar case, the consistent quantization is possible only
after including the constraint (\ref{add}).

The substitution of the field operators (\ref{or}) into the
Hamiltonian
\begin{equation}
\label{hamx} H = - \frac{1}{2} \int d\mathbf{x} \left(\pi_\mu
\pi^\mu + \mathbf{\nabla} u_\mu\mathbf{\nabla}u^\mu +m^2(t) u_\mu
u^\mu \right)
\end{equation}
gives  directly the diagonal form  in the Fock space
\begin{equation}
\label{hamp} H= -\sum\limits_{\mathbf{p}} \,\omega(\mathbf{p},t)
\, a_\mu^{*}(\mathbf{p},t)\, a_\mu (\mathbf{p},t).
\end{equation}
However, this quadratic form is not positively defined. To correct
it,  one needs to exclude the $\mu=0$ component by the additional
condition (\ref{add}). The  equations of motion for amplitudes
$a_\mu(\mathbf{p}, t)$ are similar to the scalar case
(\ref{eq_of_mot})
\begin{align} \dot{a}_\mu (\mathbf{p}, t)
= \frac{1}{2} \Delta(\mathbf{p}, t)\, a^{*}_\mu (-\mathbf{p}, t) -
i \omega(\mathbf{p}, t)\, a_\mu (\mathbf{p}, t), \label{he1}
\end{align}
where $\Delta(\mathbf{p},t)$ is defined by Eq. (\ref{Delta}).
Using Eqs. (\ref{he1}), the  condition (\ref{add}) may be
transformed now to the following relation ($i=1,2,3$)
\begin{equation}
\label{add2} \omega(\mathbf{p},t)\, a_0(\mathbf{p},t) = p_i\, a_i
(\mathbf{p},t).
\end{equation}
This equation allows one to exclude the $\mu=0$ component from the
Hamiltonian  (\ref{hamp}) and quantize other components, which
gives
\begin{equation}
\label{18} H = \sum\limits_{\mathbf{p},i,k}\, \frac{\textstyle
1}{\textstyle\omega(\mathbf{p},t)} \biggl[ \omega^2(\mathbf{p},t)
\delta_{ik} - p_i p_k \biggr] \, a_i^{\dagger}(\mathbf{p},t)\,
a_k(\mathbf{p},t),
\end{equation}
with the corresponding commutation relation and the vacuum state
\begin{equation}\label{88a}
\left[ a_i(\mathbf{p}, t), a^{\dagger}_k(\mathbf{p'}, t) \right] =
\delta_{ik} \delta_{\mathbf{pp'}} \qquad  a_i(\mathbf{p}, t\to
-\infty)|0> \,=\, 0 .
\end{equation}

The next step is the diagonalization of the quadratic form
(\ref{18}) by means of the linear transformation \cite{BSh}
\begin{equation}
\label{lin} \mathbf{a}_i(\mathbf{p},t)= E_{ik}
\alpha_k(\mathbf{p},t)  \equiv (\mathbf{e}_1)_i
\alpha_1(\mathbf{p},t) + (\mathbf{e}_2)_i \alpha_2(\mathbf{p},t)+
(\mathbf{e}_3)_i \frac{\omega}{m(t)} \alpha_3(\mathbf{p},t),
\end{equation}
where
$\{\mathbf{e}_1(\mathbf{p}),\mathbf{e}_2(\mathbf{p}),\mathbf{e}_3(\mathbf{p})\}$
is the local orthogonal basis constructed on the vector
$\mathbf{e}_3 = \mathbf{p}/|p|$. These real unit vectors form the
triad,
\begin{equation}
\label{triad} e_{ik}e_{jk} = e_{ki} e_{kj} =\delta_{ij},  \qquad
e_{ik} = (\mathbf{e}_i)_k \,.
\end{equation}
The transformation (\ref{lin}) establishes the positively-defined
Hamiltonian
\begin{equation}
\label{hd} 
H =
\sum\limits_{\mathbf{p}}\,\omega(\mathbf{p},t)\biggl[\alpha_i^{\dagger}(\mathbf{p},t)
\alpha_i(\mathbf{p},t)
+\alpha_i(\mathbf{p},t)\alpha_i^{\dagger}(\mathbf{p},t)\biggr].
\end{equation}
The presence of the $\omega/m$ factor in the non-unitary matrix
$E$ in Eq.  (\ref{lin}) leads to violation of the unitary
equivalence between the $a$-representation (\ref{18}) and
$\alpha$-representations (\ref{hd}). Equations of motion for these
new amplitudes follow from a combination of Eqs. (\ref{he1}) and
(\ref{lin})
\begin{equation}
\label{heiz_alpha} \dot{\alpha}_i (\mathbf{p},t) = \frac{1}{2}
\Delta (\mathbf{p},t) \alpha^{\dagger}_i(-\mathbf{p},t) -
i\omega(\mathbf{p},t) \alpha_i(\mathbf{p},t) +
\eta_{ij}(\mathbf{p},t)\alpha_j(\mathbf{p},t).
\end{equation}
The spin rotation matrix $\eta_{ij}$ is defined as
\begin{equation}
\label{matrix1} \eta_{ik}(\mathbf{p},t) = - \Delta_m \delta_{i3}
\delta_{k3}
\end{equation}
with $\Delta_m=-\dot{m}/m+\Delta$. This relation shows a
particular role of the third component.

Together with the Hamiltonian (\ref{hamp}), the total momentum
operator  takes also the diagonal form. However, the spin operator
\begin{equation}
\label{spin} S_i = \varepsilon_{ijk} \int d \mathbf{x}\,\biggl[
u_k \pi_j + \pi_j u_k - u_j \pi_k -\pi_k u_j\biggl]
\end{equation}
has non-diagonal terms in the spin space in terms of the operator
$\alpha_i$
\begin{equation}
\label{spin1} S_k = i \varepsilon_{ijk} \sum\limits_{\mathbf{p}}
\biggl[ \alpha_i^{\dagger}(\mathbf{p},t) \alpha_j(\mathbf{p},t) -
\alpha (\mathbf{p},t) \alpha_j^{\dagger}(\mathbf{p},t)\biggl],
\end{equation}
where $\varepsilon_{ijk}$ is the unit antisymmetric tensor. In
particular, the spin projection on the $p_3$-axis is
\begin{multline}
S_3 = i\sum\limits_{\mathbf{p}} \biggl[
\alpha_1^{\dagger}(\mathbf{p},t)\alpha_2
(\mathbf{p},t) - \alpha_2^{\dagger}(\mathbf{p},t)\alpha_1 (\mathbf{p},t) \\
+ \alpha_2 (\mathbf{p},t)\alpha_1^{\dagger}(\mathbf{p},t) -
\alpha_1(\mathbf{p},t) \alpha_2^{\dagger}(\mathbf{p},t) \biggl].
\label{s3}
\end{multline}
Thus, this representation can be named the {\em incomplete}
quasiparticle representation since the spin projection is not
fixed. The operator (\ref{s3}) can be diagonalized by a linear
transformation to the new basis \cite{BSh}
\begin{equation}
\label{lin1}
c_i (\mathbf{p},t) = R_{ik} \alpha_k (\mathbf{p},t),\\
\end{equation}
with the unitary matrix
\begin{equation}
\label{matr2} R = \frac{1}{\sqrt{2}}\left[\begin{array}{ccc}
  1 &  i & 0 \\
  -i & 1 & 0 \\
  0 & 0 & \sqrt{2} \\
\end{array}\right].
\end{equation}
As a result, the new  amplitudes $ c_i(\mathbf{p},t)$  correspond
to creation and annihilation operators of vector quasiparticles
with the total energy, 3-momentum and spin projection into the
chosen direction,
\begin{eqnarray}
\label{hamq} H(t) &=&
\sum\limits_{\mathbf{p}}\,\omega(\mathbf{p},t)\biggl[c_i^{\dagger}(\mathbf{p},t)
c_i(\mathbf{p},t) +
c_i (\mathbf{p},t) c_i^{\dagger}(\mathbf{p},t)\biggl],\qquad \\
\mathbf{\Pi}(t) &=& \sum\limits_{\mathbf{p}}
\mathbf{p}\biggl[c_i^{\dagger}(\mathbf{p},t) c_i(\mathbf{p},t) +
c_i(\mathbf{p},t) c_i^{\dagger}(\mathbf{p},t)\biggr],\\
S_3(t) &=& \sum\limits_{\mathbf{p}}\, \biggl[
c_1^{\dagger}(\mathbf{p},t) c_1(\mathbf{p},t) -
c_1(\mathbf{p},t) c_1^{\dagger}(\mathbf{p},t) \nonumber\\
&& \qquad + c_2(\mathbf{p},t) c_2^{\dagger}(\mathbf{p},t)-
c_2^{\dagger}(\mathbf{p},t) c_2(\mathbf{p},t) \biggl].
\end{eqnarray}
This $c$-representation will be referred to as  the {\em complete}
quasiparticle representation. The equations of  motion for these
amplitudes follow from Eqs. (\ref{heiz_alpha}),(\ref{lin1})
\begin{equation}
\label{heisc} \dot{c}_i (\mathbf{p},t) =  \frac{1}{2} \Delta
(\mathbf{p},t) c_i^{\dagger}(-\mathbf{p},t) -
i\omega(\mathbf{p},t) c_i(\mathbf{p},t) +
\eta_{ij}(\mathbf{p},t)c_j(\mathbf{p},t),
\end{equation}
where the matrix $\eta_{ij}$ is fixed by Eq.(\ref{matrix1}).

The transition to this representation from the initial
$a$-representation is defined by the combination of
transformations (\ref{lin}) and (\ref{lin1})
\begin{equation}
\label{lin3} {c}(\mathbf{p},t) = U(\mathbf{p},t) {a}(\mathbf{p},t)
\end{equation}
with the nonunitary operator ($\mathbf{e}^{(\pm)} =
(\mathbf{e}_1\pm i\mathbf{e}_2)/\sqrt{2}$)
\begin{equation}
\label{lin_u} U(\mathbf{p},t) = R \cdot E^{-1}(\mathbf{p},t) =
    \begin{bmatrix}
        e_1^{(+)} & e_2^{(+)} & e_3^{(+)}\\
        e_1^{(-)} & e_2^{(-)} & e_3^{(-)}\\
        \frac{m}{\omega}e_{31} & \frac{m}{\omega}e_{32} & \frac{m}{\omega}e_{33}
    \end{bmatrix}.
\end{equation}

To solve the quantization problem,   the equation of motion should
be taken into account. The commutation relation has the
non-canonical form
\begin{equation} \label{cr}
\bigl[c_i(\mathbf{p},t),c^{\,+}_j(\mathbf{p'},t)\bigr] =
Q_{ik}(\mathbf{p},t)Q_{jk}(\mathbf{p},t)
\delta_{\mathbf{p}\mathbf{p}'},
\end{equation}
where the matrices $Q_{il}(\mathbf{p},t)$ are defined by the
equations
\begin{equation}
\label{qq} \dot{Q}_{ij}(\mathbf{p},t) =
\eta_{ik}(\mathbf{p},t)Q_{kj}(\mathbf{p},t)
\end{equation}
with the initial conditions
\begin{equation} \label{qic} \lim_{t\rightarrow -\infty}{Q}_{ij}(\mathbf{p},t) =
\delta_{ij}. \end{equation}
So the commutation relation is
transformed to the canonical form only in the asymptotic limit
$t\rightarrow -\infty$. Relation (\ref{cr}) provides the
definition of positive-energy quasiparticle excitations of vacuum
to be treated as some time-dependent energy reservoir .

\subsection{Kinetic equations \label{Sec_KE}}

The standard procedure to derive the KE \cite{Schmidt} is based on
the Heisenberg-type equations of motion (\ref{he1}) or
(\ref{heisc}). Let us introduce  one-particle correlation
functions of vector bosons in the initial $a$-representation
\begin{eqnarray} \label{distr}
F_{ik}(\mathbf{p},t) & = & \langle 0|a^{\dagger}_i (\mathbf{p},t)
a_k (\mathbf{p},t)|0\rangle,
\end{eqnarray}
where the averaging procedure is performed over the in-vacuum
state \cite{Grib}. By differentiating the first equation with
respect to time, we obtain
\begin{equation}
\label{36} \dot{F}_{ik}(\mathbf{p},t) =  \frac12 \Delta
(\mathbf{p},t) \left\{ \Phi_{ik}(\mathbf{p},t) +
\Phi^{\dagger}_{ik}(\mathbf{p},t) \right\},
\end{equation}
where the auxiliary correlation function is introduced as
\begin{equation}
\label{anom} \Phi_{ik}(\mathbf{p},t) = \langle 0|a_i
(-\mathbf{p},t) a_k (\mathbf{p},t)|0\rangle.
\end{equation}
Equation of motion for this function can be obtained by analogy
with Eq. (\ref{36}). We write out the answer in the integral form
\begin{equation}
\label{38} \Phi_{ik}(\mathbf{p},t) = \frac{1}{2}
\int\limits_{-\infty}^{t} dt' \Delta(\mathbf{p},t')
\left[\delta_{ik} + 2 F_{ik}(\mathbf{p},t') \right] e^{
2i\theta(\mathbf{p};t,t')}.
\end{equation}
The asymptotic condition $F_{ik}(\mathbf{p},-\infty) = 0$ (the
absence of quasiparticles in the initial time) has been used here.
The substitution of Eq. (\ref{38}) into Eq. (\ref{36}) leads to
the resulting KE
\begin{multline}
\label{ke} \dot{F}_{ik}(\mathbf{p},t) = \frac{1}{2} \Delta
(\mathbf{p},t) \int\limits_{-\infty}^{t} dt' \Delta
(\mathbf{p},t') [\delta_{ik} + 2 F_{ik}(\mathbf{p},t')] \cos[ 2\,
\theta (\mathbf{p};t,t') ].
\end{multline}
This KE is a natural generalization of the corresponding KE for
scalar particles (Sec. \ref{KinEq}).

However, there is a number of problems that are specific for the
theory of massive bosons: The energy is not positively-defined,
the spin operator has non-diagonal terms in the space of spin
states {\em etc}. This circumstance hampers the physical
interpretation of the distribution function (\ref{distr}). In
order to overcome this difficulty, it is necessary to proceed to
the complete QPR where the system has well-defined values of
energy, spin etc. The simplest way to derive the KE in this QPR is
based on the application of the transformation (\ref{lin3})
directly to the KE (\ref{ke}).

Similarly to the definitions (\ref{distr}), we introduce
correlation functions of vector particles in the complete QPR
\begin{eqnarray}
\label{distr1} f_{ik}(\mathbf{p},t) & = & \langle 0|c^{\dagger}_i
(\mathbf{p},t) c_k (\mathbf{p},t)|0\rangle.
\end{eqnarray}
They are connected with the primordial  correlation functions
(\ref{distr}) by relations of the type
\begin{equation}
\label{dfdf}
f_{ik}(\mathbf{p},t)=U^{\dagger}_{in}(\mathbf{p},t)U_{km}(\mathbf{p},t)
F_{nm}(\mathbf{p},t),
\end{equation}
As a result, the KE (\ref{ke}) becomes:
\begin{multline} \label{keq1}
\dot{f}_{ik}(\mathbf{p},t) = \frac{1}{2} \Delta(\mathbf{p},t)
\int\limits^t_{-\infty} dt'
\Delta(\mathbf{p},t')M_{ikjl}(\mathbf{p},t,t') \bigl[\,
 \delta_{jl}  \\ + 2 f_{jl}(\mathbf{p},t')
 \bigr] \cos{2\theta(\mathbf{p};t,t')} - \Delta_m(\mathbf{p},t) [
\delta_{i3}f_{3k}(\mathbf{p},t) + \delta_{k3}f_{i3}(\mathbf{p},t)
],
\end{multline}
where
\begin{multline}
M_{ikjl}(t,t') = \delta_{ij}^\perp \delta_{kl}^\perp +
\frac{\omega(t')}{\omega(t)}\frac{m(t)}{m(t')}\left[
\delta_{i3}\delta_{j3} \delta^\perp_{kl} + \delta_{k3}\delta_{l3}
\delta^\perp_{ij} \right. \\ + \left.
\frac{\omega(t')}{\omega(t)}\frac{m(t)}{m(t')}
\delta_{i3}\delta_{k3}\delta_{j3}\delta_{l3}\right]
\end{multline}
and $\delta_{ik}^\perp =\delta_{ik}-\delta_{i3}\delta_{k3}$.

As was expected, distribution functions
$f_{\alpha\beta}(\mathbf{p},t)$  satisfy the same KE (\ref{ke})
for $\alpha=1,2$. The feature of the complete QPR becomes apparent
only in tensor components of the distribution function
$f_{ik}(\mathbf{p},t)$ which contains the preferred values of the
spin index $i$ and (or) $k=3$. Let us select the KE for diagonal
components of the correlation function (\ref{distr1}) which has
the direct physical meaning of transversal ($i=1,2$)
\begin{eqnarray}
\dot{f}_{i}(\mathbf{p},t) = \frac12 \Delta(\mathbf{p},t)
\int\limits^t_{-\infty} dt' \Delta(\mathbf{p},t') \bigl[1+ 2
f_{i}(\mathbf{p},t')\bigr] \cos{2\theta(\mathbf{p};t,t')}
\label{kema}
\end{eqnarray}
and longitudinal components of the distribution function
\begin{multline}\dot{f}_{3}(\mathbf{p},t) = -2\Delta_m(\mathbf{p},t)
f_{3}(\mathbf{p},t)\, + \frac12 \Delta(\mathbf{p},t)
\frac{m^2(t)}{\omega^2(t)}\\ \times \int\limits^t_{-\infty} dt'
\Delta(\mathbf{p},t') \frac{\omega^2(t')}{m^2(t')}
\bigl[2f_{3}(\mathbf{p},t') + Q(\mathbf{p},t')\bigr]
\cos{2\theta(\mathbf{p};t,t')}. \label{kem3}
\end{multline}
Here the shorthand notation $f_{ii}=f_i$ has been introduced for
diagonal components of the matrix correlation functions
(\ref{distr1}) and
\begin{eqnarray}
\label{dd} \Delta =\frac{m \dot{m}}{\omega^2}, \qquad \Delta_m =
-\Delta \frac{\mathbf{p}^2}{m^2}.
\end{eqnarray}
Longitudinal and transversal distribution functions are connected
by the relation
\begin{equation}
\label{48} f_{3}(\mathbf{p},t) = Q
(\mathbf{p},t)f_{1}(\mathbf{p},t),
\end{equation}
where  $Q (\mathbf{p},t)$ is the function entering into the
commutation relation for the longitudinal bosons,
\begin{gather}
\label{49} [c_3(\mathbf{p},t), c_3^{\dagger}(\mathbf{p'},t)] =
Q(\mathbf{p},t)\,
\delta_ {pp'}, \nonumber \\
Q (\mathbf{p},t) = \exp{\biggl[-2\int_{t_0}^t \Delta_m(t')
dt'\biggr]} =
\left[\frac{m(t)}{m(t_0)}\frac{\omega(t_0)}{\omega(t)}\right]^2
\end{gather}
with the  corresponding initial values $m(t_0)$ and $\omega
(t_0)$.

Due to relation (\ref{48}), it is sufficient to solve only Eq.
(\ref{kema}). The KE for transversal bosons is transformed from
the integro-differential form to a set of ordinary differential
equations similarly to the cases considered above (Sects.
\ref{KinEq} and \ref{secf2}):
\begin{equation}
\label{58} \dot{f}_k = \frac{1}{2} \Delta u_k,\qquad \dot{u}_k =
\Delta  (1 + 2{f}_k) - 2\omega v_k,\qquad \dot{v}_k =2\omega u_k.
\end{equation}
The general initial condition for all diagonal components of  the
distribution function
\begin{equation}
\label{ic} \lim_{t \rightarrow - \infty} f_k(t) = \lim_{t
\rightarrow - \infty} u_k(t)  = \lim_{t \rightarrow - \infty}
v_k(t) = 0
\end{equation}
satisfies the following requirement
\begin{equation}
\label{icm} \lim_{t \rightarrow - \infty} m(t) = m_0,  \quad
\mbox{or}  \quad \lim_{t \rightarrow - \infty} \dot{m}(t) = 0.
\end{equation}

The main operating characteristic of the vacuum creation process
is the total number  density of vector  bosons,
\begin{eqnarray}\label{densv} n_{tot}(t) &=& \frac{1}{\pi^2}
\int\limits^\infty_0 p^2 dp\, \bigl[2f_1(p,t) + f_3(p,t)\bigr] \nonumber \\
&=& \frac{1}{\pi^2} \int\limits^\infty_0 p^2 dp\, f_1(p,t) \bigl[
2+Q(p,t)\bigr]~, \end{eqnarray}
where isotropy of the system was assumed, $p=|\mathbf{p}|$. As it
will be shown in Sec. \ref{EOS}, the integral (\ref{densv}) is
convergent. It essentially differs from results based on the
Wentzel Lagrangian (\ref{lagr_w}). In particular, as was shown in
\cite{DB,Perv02,BPZZ06}, the Wentzel approach leads to infinite
particle density. Authors of these papers give some arguments in
favor of that allowing for quasparticle collisions will result in
finite physical quantities and bring the estimated energy budget
of the Universe to agreement with observation data
\cite{Perv02,GPVZZ04}.

We will consider below the case when the time dependence of the
vector boson mass is defined by conformal evolution of the
Universe. The set of equations similar to (\ref{58}) was obtained
first in \cite{Ver} within an alternative model Lagrangian
(\ref{lagr_w}) in the Friedman-Robertson-Walker (FRW) space-time.
The quantization procedure is quite ordinary in this approach. It
would be useful to compare the predictions of these two models in
detail.

\subsection{EoS for the isotropic case \label{EOS}}

The relations for the energy density and pressure can be derived
from the energy-momentum tensor corresponding to the Lagrangian
(\ref{Lagr_dens})
\begin{equation} \label{e1} T_{\mu\nu} = -\partial_{\mu}u_{\alpha}
\partial_{\nu}u^{\alpha} - \partial_{\nu}u_{\alpha}
\partial_{\mu}u^{\alpha} - g_{\mu\nu}\mathcal{L}. \end{equation}
Taking into account the isotropy of the system, we obtain the EoS
for  the massive vector boson gas \cite{DB}
\begin{eqnarray}
\label{e6}
\varepsilon(t) &=& 2 \int [dp]\, \omega ( 2 + Q ) f_1, \nonumber \\
P (t) &=& \frac{2}{3}\varepsilon(t) - \frac{4}{3} m^2 \int
\frac{[dp] }{\omega} ( 2 + Q ) f_1  + \delta P_{vac}
(t)\label{e7},
\end{eqnarray}
where $\delta P_{vac} (t)$ is the contribution to pressure induced
by  the vacuum polarization,
\begin{equation}
\label{e7a} \delta P_{vac} (t) = -  \frac{2}{3} \int \frac{[dp]
}{\omega}\, \biggl(2\omega^2 + m^2 \biggr) \left[ 1+ Q
\left(\frac{1}{2}+ \frac{\mathbf{p}^2}{m^2}\right)\right] u_1.
\end{equation}

In order to prove the convergence of the integrals (\ref{e6}),
(\ref{e7a}) we investigate the asymptotic behavior of the solution
of the system of equations (\ref{58}). This system can be solved
exactly in the limit $p\gg m$ for the case $\alpha=1/2$ (the
parameter $\alpha$ is defined in the Sec. 5)
\begin{eqnarray}
\label{p1} \dot{f} = \frac{m_H^2}{4t_H}\frac{1}{ p^2} u,\qquad
\dot{u} = \frac{m_H^2}{2t_H}\frac{1}{ p^2}(1+2f) - 2p v, \qquad
\dot{v} = 2 p u~.
\end{eqnarray}
The asymptotic solution of Eqs. (\ref{p1}) with the initial
conditions (\ref{ic}) is
\begin{align}\label{p2}
f(p,t) = \frac{v(p,t)}{(2p/m_0)^3}  \sim
\frac{\sin^2{p(t-t_0)}}{16 (p/m_0)^6}, \qquad u(p,t) \approx
\frac{\sin{2 p(t-t_0)}}{4 (p/m_0)^3},
\end{align}
where $m_0=m(t_0)=m_H^{2/3}t_H^{-1/3}$ and $t_0 = 1/m_0$. The
numerical study of Eqs. (\ref{58}) shows that the basic features
of the solutions (\ref{p2}) for $\alpha = 1/2$ are conserved also
for other $\alpha > 0$. It corresponds to results of Sect.
\ref{SF3}.

According to Eq. (\ref{p2}), the particle number density and
energy density (see Eqs. (\ref{densv}) and (\ref{e6}),
respectively) are convergent, but the vacuum polarization
contribution to pressure (\ref{e7a}) is divergent. Moreover,
irrelevant fast vacuum oscillations of the pressure are observed
here. Such a behavior of the pressure for a plasma created from
vacuum is not specific for the present theory but it is inherent
in the models where an electron-positron plasma is created in
strong time-dependent electric fields as investigated in
\cite{23}. The standard regularization procedure of similar
integrals with some unknown functions satisfying ordinary
differential equations is based on the study of asymptotic
decompositions of these functions in power series of the inverse
momentum, $1/{p}^N$ (see Sect. \ref{SF3}). In the considered case,
such a procedure is not effective because the solution (\ref{p1})
has quickly oscillating factors ("Zitterbewegung"\,), whose
asymptotic decomposition leads to secular terms. Therefore, for
numerical calculations we regularize the pressure by a momentum
cut-off at $p=10m_0$  and separate its stable part by means of the
time averaging procedure
\begin{equation}
 \label{3} <P > = \frac{1}{(t-t_0)}\int\limits_{t_0}^t p(t) dt.
\end{equation}
Such "coarse graining"\, procedure was proposed in \cite{19} in
order  to exclude the "Zitterbewegung"\, from the description of
vacuum particle creation. In reality,  these fast oscillations are
smoothed out due to dissipative processes which are not taken into
consideration here (the first attempt to derive the collision
integral for the scalar quasiparticle gas in a strong electric
field was made in \cite{CI}).

\section{Application to conformal cosmology models \label{UNI}}

The description of the vacuum creation of particles in
time-dependent gravitational fields of cosmological models goes
back to Refs. \cite{Parker,GM,SU,Z} and has been reviewed, {\em
e.g.}, in monographs \cite{Birrell,ZN,Linde}. The particularity of
our work consists in the consideration of vacuum generation of
particles at  conditions of the early Universe in the framework of
a conformal-invariant cosmological model \cite{DB,Perv02}. Thus,
the space-time is assumed to be conformably flat and  the
expansion of the Universe in the Einstein frame (with metric
$\tilde{g}_{\mu\nu}$) with constant masses $\tilde{m}$ can be
replaced by the change of masses in the Jordan frame (with metric
$g_{\mu\nu}$) due to the evolution of the cosmological (scalar)
dilaton background field \cite{FM,BD}. This mass change is defined
by the conformal factor $\Omega(x)$ of the conformal
transformation
\begin{equation}
\tilde{g}_{\mu\nu}(x) = \Omega^2 (x)g_{\mu\nu}.
\end{equation}
Since mass terms generally violate the conformal invariance, a
space-time dependent mass term
\begin{equation}
{m}(x) = \frac{1}{\Omega(x)}\tilde{m}
\end{equation}
has been introduced which formally keeps the conformal invariance
of the theory \cite{Birrell}. In the important particular case of
the isotropic FRW space-time, the conformal factor is equal to the
scale factor, $\Omega(x)= a(\tilde{t})$, and hence $m(\tilde{t}) =
a(\tilde{t})m_{obs}$, where $\tilde{t}$ is "the Einstein time"\,
and $m_{obs}$ is the observable present-day mass. Such a
dependence was used, {\em e.g.}, in Ref. \cite{Grib} for the FRW
metric. On the other hand, the scaling factor $a(\tilde{t})$ is
defined by the cosmic equation of state. For a barotropic fluid,
this EoS has the form
\begin{equation}
P_{ph} = (\gamma - 1)\varepsilon_{ph} = c^2_s \varepsilon_{ph},
\end{equation}
where $P_{ph}$ and $\varepsilon_{ph}$ are phenomenological
pressure and energy density (in contrast to "dynamical"\, $P$ and
$\varepsilon$, see Sect. \ref{EOS}), $\gamma$ is the barotropic
parameter, $c_s$ is the sound velocity. The solution of the
Friedman equation for such EoS leads to the following scaling
factor:
\begin{equation}
\label{scaling}
a(\tilde{t}) \sim \tilde{t}^{\ 2/3\gamma}.
\end{equation}
Kinetics of the vacuum creation of massive vector bosons (Sect.
\ref{Sec_KE}) was constructed in the flat Jordan frame with the
proper conformal time ${t}$ which is necessary to introduce now in
Eq. (\ref{scaling}). The transition to the conformal time is
defined by the relation $dt = d\tilde{t}/a(\tilde{t})$. From this
relation and Eq. (\ref{scaling}) it follows
\begin{equation}
\label{time_scaling} \tilde{t} \sim \left[\left(1 -
\frac{2}{3\gamma}\right)\,  t\right]^{3\gamma / (3\gamma - 2)}.
\end{equation}
The substitution of this relation into Eq. (\ref{scaling})
establishes the mass evolution law in the terms of the conformal
time
\begin{equation}
\label{mass} m(t) = (t/t_H ) ^{\alpha}\, m_W, \qquad  \alpha =
\frac{2}{3\gamma - 2},
\end{equation}
where $t_H = [(1+\alpha)H]^{-1}$ is the scaling factor (the age of
the Universe),  $H=70~km/s/Mpc$ is the the present-day Hubble
constant and the W-boson mass is taken as $m_W = 80$~GeV.  Values
of the $\alpha$ parameter for some popular EoS are: $\gamma =2$,
$\alpha = 1/2 $ (stiff fluid); $\gamma = 4/3$,~$\alpha = 1$
(radiation); $\gamma =1$,~$\alpha = 2$ (dust);  $\gamma < 2/3$
(quintessence); $\gamma = 0,~\alpha = -1$ (cold matter including
baryon mass and dark matter) \cite{Dolgov,Ger06}.
\begin{figure}[t]
\centering
\includegraphics[width=0.48\textwidth,keepaspectratio]{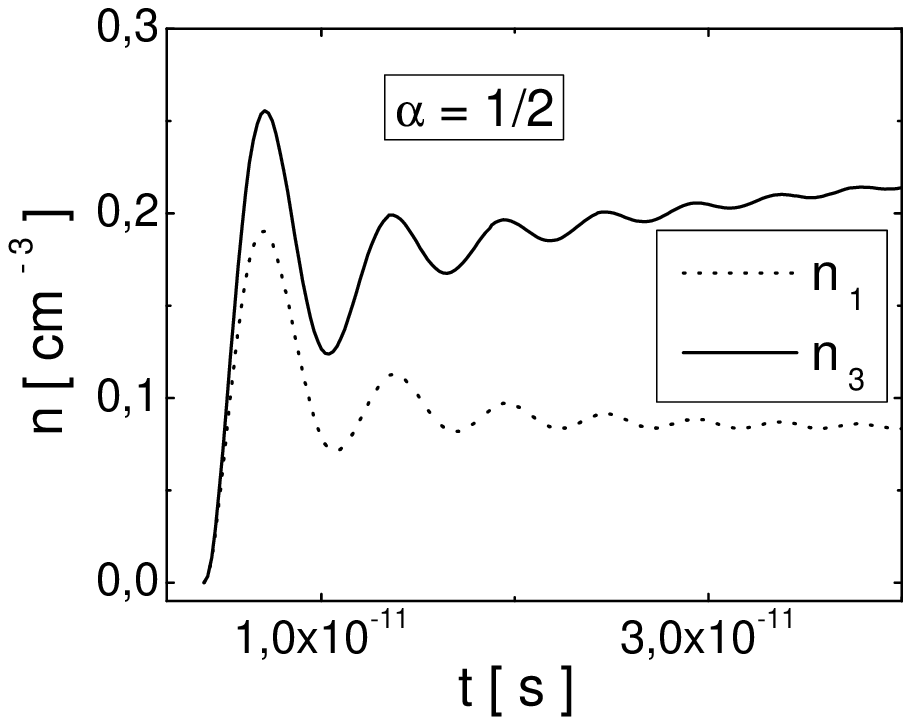}
\includegraphics[width=0.48\textwidth,keepaspectratio]{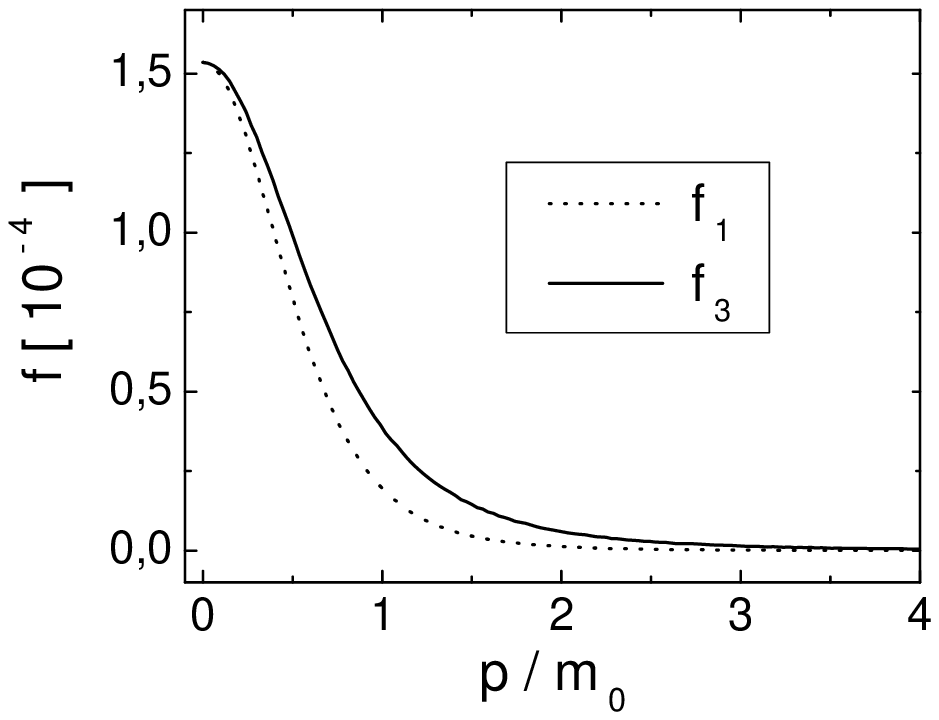}
\caption{Time evolution of the number density for transversal
$n_1$ and longitudinal $n_2$ vector bosons with the initial
condition $m_0 \cdot t_0 = 1 $ for $\alpha=1/2$ (left panel) and
the corresponding momentum distribution at the time $t \gg t_0$
(right panel).} \label{time}
\end{figure}
\begin{figure}[t]
\centering
\includegraphics[width=0.48\textwidth,keepaspectratio]{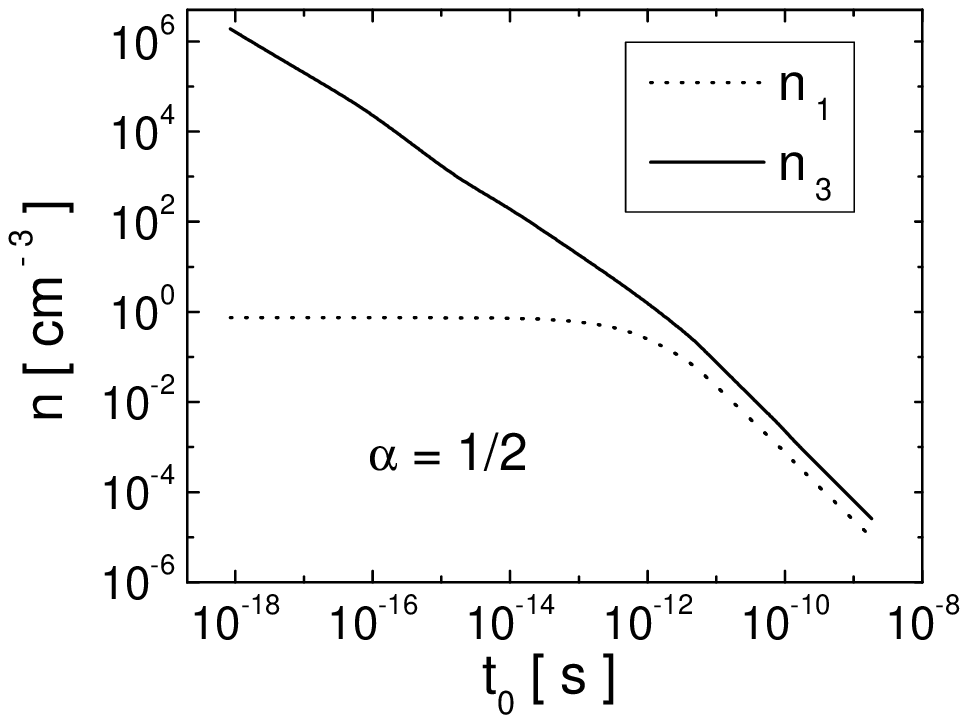}
\includegraphics[width=0.48\textwidth,keepaspectratio]{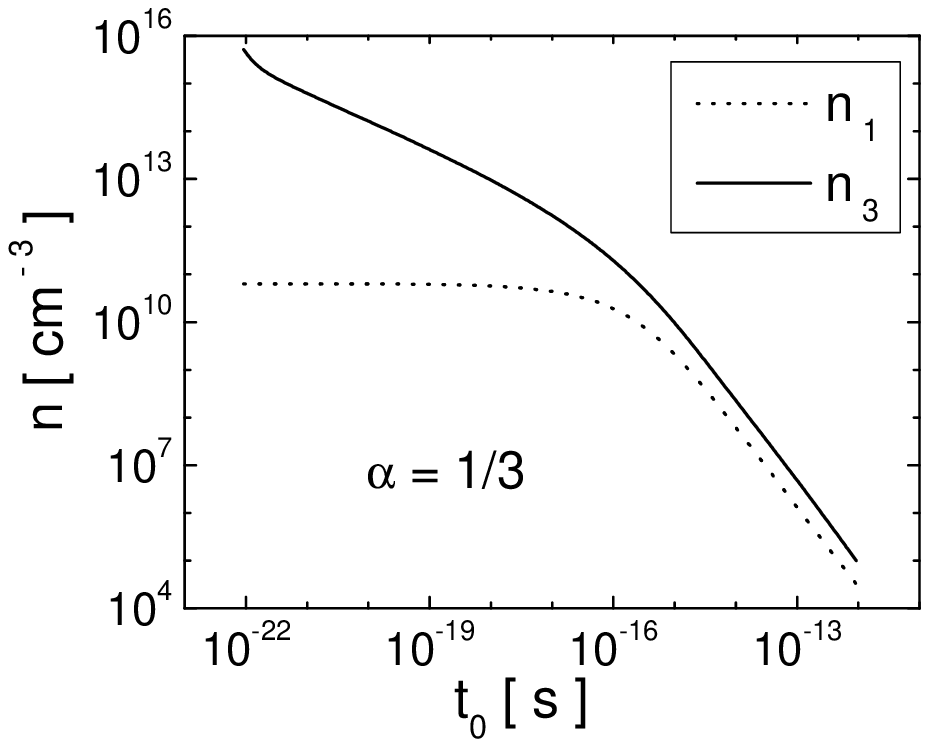}
\caption{The dependence of the number density of residual vector
bosons on the initial time $t_0$:   $\alpha=1/2$ (left panel),
$\alpha=1/3$ (right panel).} \label{final}
\end{figure}

Due to back reactions and the dynamical mass generation during the
cosmic  evolution the detailed mass history remains to be worked
out. The central question, however, is whether the number density
of produced W-bosons could be of the same order as that of the
cosmic microwave background (CMB) photons, $n_{\rm CMB}\sim
465~{\rm cm}^{-3}$. If this question may be answered positively,
the vacuum pair creation of W-bosons from a time-dependent scalar
field (mass term) could be suggested as a mechanism for the
generation of matter and radiation in the early Universe.

The numerical analysis of Eqs. (\ref{58}) for massive vector
bosons is performed by the standard Runge-Kutta method on a
one-dimensional momentum grid. As one can see in Fig. \ref{time},
the  creation process ends very quickly and the particle density
saturates at some finite value. The momentum distribution of
particles is formed also very early when $m(t)\approx m_0$ and
frozen in such form so later on, for times $t \gg t_0$,  most of
particles have very small momenta $p \ll m(t)$. The spectrum of
created  bosons is essentially nonequilibrium; hence we should
continue  the analysis of relevant dissipative mechanisms and
other observable manifestations of a non-equilibrium state (e.g.,
CMB photons; in this connection, see, for example,
\cite{Komatsu,VMD}).

The dependence of the corresponding final value of density on the
initial time is shown in Fig. \ref{final}. The final density $n_1$
of the transversal vector bosons with spin projection $\pm 1$
reaches a maximum when for very early initial times we are close
to the birth of the Universe. However, in the same limit, the
density $f_3$ of the longitudinal particles with zero spin
projection
 grows beyond all bounds. The choice of the EoS changes
drastically the number of created particles, thus resulting in
values which are too small ($\alpha = 1/2$) or too large ($\alpha
= 1/3$) in comparison with the observed CMB photon densities. In
order to improve this model, we should use an improved EoS,
assuming that the barotropic parameter $\gamma$ characterizing the
evolution of particle masses can be changed during the time
evolution. Such a time-dependence could be induced by action of
the back-reaction of created particles on the scalar field.
Furthermore, we could use another space-time model, {\em e.g.},
the Kasner space-time \cite{SmolHP} instead of the conformally
flat de Sitter one. As compared to the earlier work \cite{DB}, the
main achievement  of this approach  is that there is no divergence
in the distribution function; thus we do not need to introduce any
ambiguous regularization procedure.

\begin{figure}[t]
\centering
\includegraphics[width=0.48\textwidth,keepaspectratio]{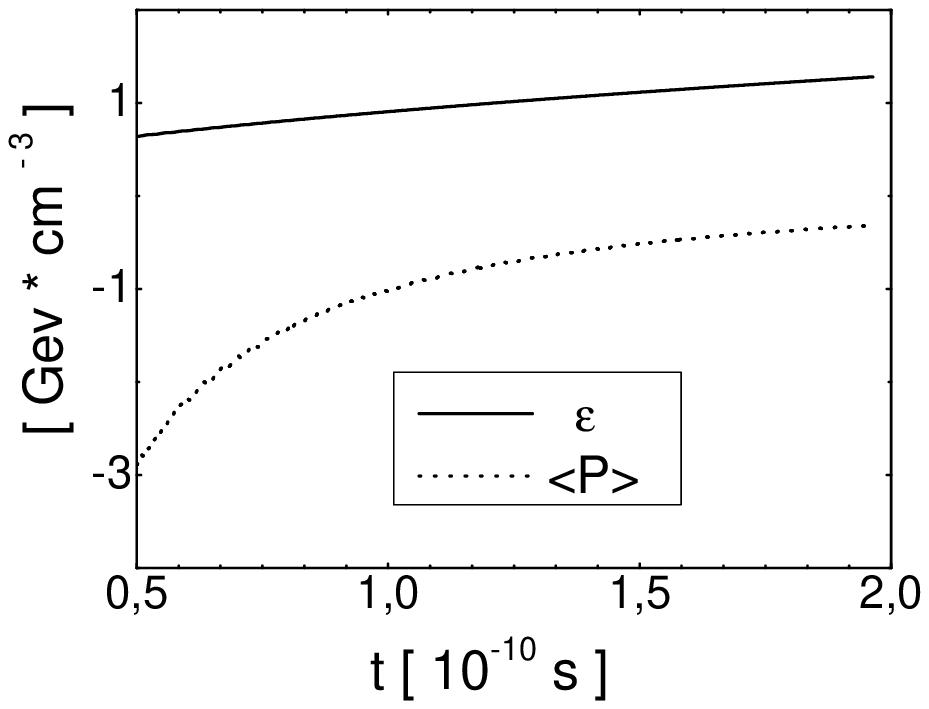}
\includegraphics[width=0.48\textwidth,keepaspectratio]{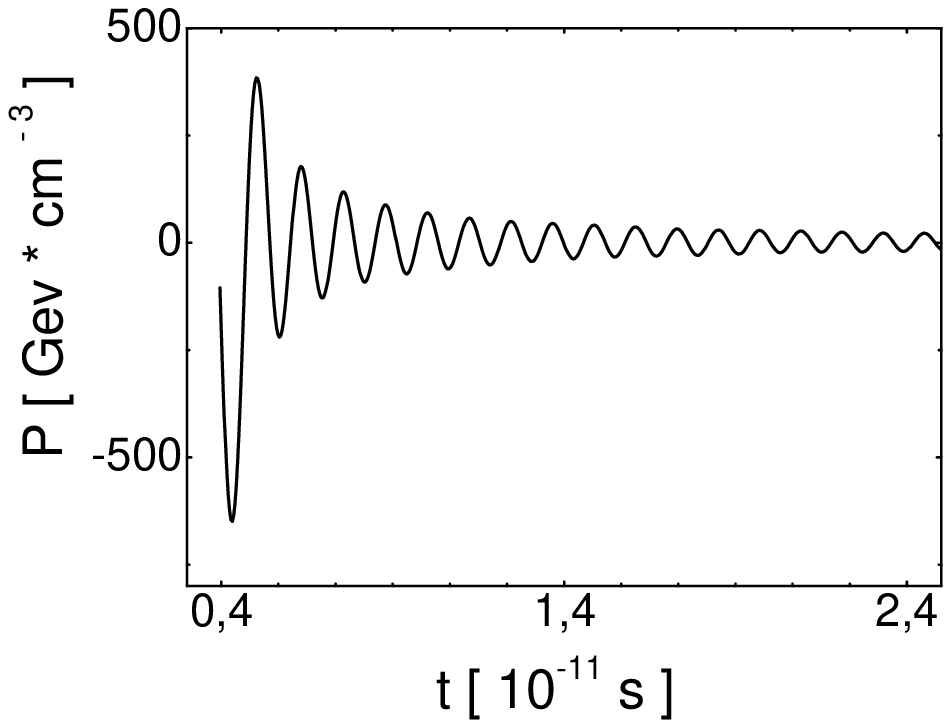}
\caption{The time dependence of the  energy density and the mean
value of pressure (\ref{3}) at $t\gg t_0$ (left panel) as well as
pressure (\ref{e6}) (right panel)  with the initial condition
$m_0\cdot t_0=1$ for $\alpha=1/2$.} \label{press}
\end{figure}

\begin{figure}[t]
\centering
\includegraphics[width=0.48\textwidth,keepaspectratio]{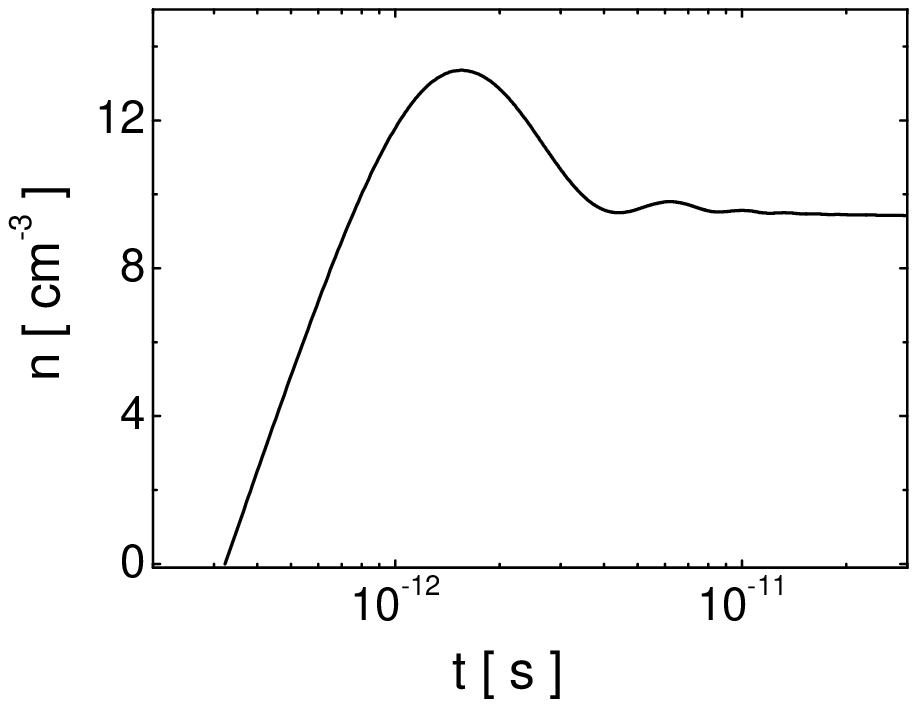}
\includegraphics[width=0.48\textwidth,keepaspectratio]{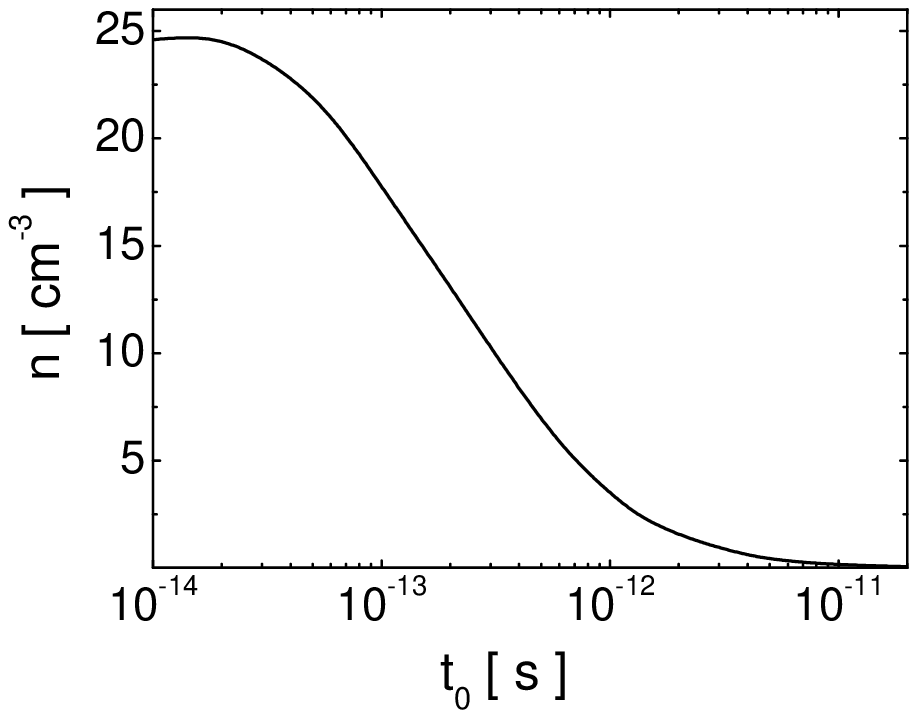}
\parbox[t]{0.48\textwidth}{
\caption{The time dependence of fermion pair density for a mass
evolution with $\alpha = 1/2$ and $m_q=170$ GeV.}\label{dens-t}}
\hspace{2.5mm}
\parbox[t]{0.48\textwidth}{
\caption{The final density of fermions for the conformal time
dependence of mass (\ref{mass}) as a function of the initial time
$t_0$ for $\alpha = 1/2$.} \label{dens_conf}}
\end{figure}

As shown in Fig. \ref{press},  the mean pressure remains negative
and its magnitude becomes negligible in comparison with the energy
density (as to the role of the negative pressure in cosmology see,
{\em e.g.}, \cite{Rama}). These features can lead to violation of
the energy dominance condition $\epsilon + P \geq 0 $ that
corresponds to accelerated expansion of the Universe. Such a kind
of models is widely  discussed (e.g., \cite{Rubakov}). For large
time moments, the energy density grows but the pressure stays very
small, $ P \simeq 0$. The energy growth under condition $ P \simeq
0$ results in the conclusion that the massive vector
boson-anti-boson gas created from the vacuum is cold.  It can be
seen directly from Eq. (\ref{e6}) that at large times $\varepsilon
(t) \simeq m(t)n_{tot}(t)$, because of $\omega (t) \simeq m(t)$.
This EoS of the massive vector boson gas ($\varepsilon \neq 0$ and
$p \simeq 0$) corresponds to dust-like matter \cite{20}, which
would characterize the evolution of the Universe  when the vector
boson gas is the dominant component of its matter (energy)
content. At a qualitative level, this conclusion is valid
independently of the specific choice of the EoS and, in
particular, in the case of the dust-like EoS. It would be
interesting to obtain a formula like (\ref{mass}) as a result of
the solution of the Friedman equation with the EoS
(\ref{e6})-(\ref{e7a}) (such a procedure represents the back
reaction problem) and investigate self-consistently the production
of vector bosons in the Universe. Let us remark that vacuum
creation of massive vector bosons in the FRW metric was first
considered in \cite{Ver}.

In order to investigate features of the fermion vacuum creation in
the conformal model of the Universe, let us use the corresponding
basic KE (\ref{FermKE}) and, as an example, choose the system of
heavy top quarks with the mass $m_q=170$~GeV (with the change
$m_w\to m_q$ in Eq. (\ref{mass})). The values of the parameter
$\alpha$ are the same as  for the vector boson. As one can see in
Fig. \ref{dens-t}, the  creation process is completed very quickly
and the particle density saturates at some finite value.

Finally, Fig. \ref{dens_conf} shows the dependence of the residual
density on the initial time $t_0$ for $m(t)$ given by Eq.
(\ref{mass}). At the qualitative level, the same picture will take
place for the vacuum creation of neutralino, which can be the main
component of dark matter (e.g., \cite{Nevzorov}). An analogous
problem in early cosmology was considered in \cite{Prokopec}.

\section{Systems with metastable vacuum \label{meta}}
\subsection{Formulation of the problem \label{meta1}}

Here a rather general mechanism of the mass formation  as a result
of  self-consistent dynamics of mean-field and quantum
fluctuations  will be considered. The separation of the
quasiclassical background field is a common procedure of different
non-perturbative approaches in QFT
\cite{Grib,Birrell,Kluger,Shur}. In the framework of this
procedure quantum fluctuations can be described by the
perturbation theory.

There is a class of physics problems in which the  strong
background field creates particles which in turn influence the
background field (the back reaction problem). In this respect it
is worthy to mention such problems as  the decay of disoriented
chiral condensate \cite{Keiser}, the resonant decay of CP-odd
metastable states \cite{Blaschke,Ahrensmeier}, the pre-equilibrium
QGP evolution \cite{Schmidt,Schmidt2,Schm4,Schm3}, the phase
transition in systems with the broken symmetry \cite{Baacke} {\em
etc}.

The construction of general kinetic theory of such a kind for
various potentials is presented in Sect. \ref{meta2}. We will
derive the closed system of equations for the self-consistent
description of the back reaction problem, including  the KE with
nonperturbative source term describing the particle creation in
the quasi-classical background field and the equation of motion
for this background field. We use the OR to derive  the KE. As an
illustrative examples, in Sect.~\ref{meta3} the one-component
scalar theory with $\Phi^4$ and double-well potential are
considered. In these examples, we study some features of the
proposed approach. In particular, the problem of the stable vacuum
state definition and possibility to emerge tachyonic regimes is
discussed. As a less trivial example, the pseudo-scalar sector of
the Witten-De Vecchia-Veneziano model will be considered.  Similar
analysis was carried out in some other models of such kind (e.g.,
\cite{Blaschke, Ahrensmeier,FL,Loh,Hama}). In this section, we
follow paper \cite{PL05}.

\subsection{The set of basic equations \label{meta2}}

Let us consider now the   scalar field Lagrangian with   a
self-interaction potential $V[\Phi]$
\begin{eqnarray} \label{sm:lag1}
\mathcal{L}[\Phi]&=&\frac{1}{2}\partial_\mu \Phi \ \partial^\mu
\Phi- \frac{1}{2}m_0^2 \Phi^2 - V[\Phi], \label{sm:sm_lag2}
\end{eqnarray}
 where $m_0$ is the bare mass and the potential $V[\Phi]$ is an
arbitrary continuous function with at least one minimum that is
necessary for a correct definition of the vacuum state. It is
assumed that the field $\Phi$ may be decomposed into the
quasiclassical space-homogeneous time-dependent background field
$\phi_0(t)$ and fluctuation part $\phi(x)$
\begin{equation}
\label{sm:back} \Phi(x)=\phi_0(t)+\phi(x).
\end{equation}
In accordance with the definition of fluctuations, we have
$\langle\phi\rangle=0$ and $\langle\Phi\rangle = \phi_0$, where
the symbol $\langle\ldots\rangle$ denotes some averaging
procedure. The background field $\phi_0(t)$ can be treated as
quasi-classical one at the condition \cite{BLP}
\begin{equation}\label{sm:quasi}
  |\dot{\phi}_0| \gg 1/(\delta t)^2 , 
\end{equation}
where $\delta t$ is the characteristic time of the field
averaging.

We consider  the case of quite small fluctuations in the vicinity
of the background field. Therefore,  the potential energy
expansion in powers of $\phi(x)$ can be performed
\begin{eqnarray}
\label{sm:dec} V[\Phi] = V[\phi_0] + R_1 \phi + \frac{1}{2} R_2
\phi^2 + V_r[\phi_0, \phi],
\end{eqnarray}
where
 \begin{eqnarray}\label{sm:6}
 R_1 = R_1[\phi_0] = \frac{d V[\phi_0]}{d \phi_0},\qquad
  R_2 = R_2[\phi_0] =  \frac{d^2 V[\phi_0]}{d \phi_0^2}
\end{eqnarray}
and  $V_r[\phi_0, \phi]$ is a residual term containing the higher
order contributions to be neglected in the current consideration
(non-dissipative approximation). The decomposition (\ref{sm:dec})
can be finite (for polynomial theories) or infinite.  After field
decomposition (\ref{sm:back}) the equation of motion
\begin{equation}\label{sm:mot_g}
 \partial_\mu\partial^\mu  \Phi  + m_0^2 \Phi  +
\frac{d V[\Phi]}{d \Phi} = 0
\end{equation}
can be rewritten in  the following form:
\begin{equation} \label{sm:mot}
 (-\partial_\mu\partial^\mu - m^2)\phi = Q[\phi_0, \phi],
\end{equation}
 where the relation
 \begin{equation}
\label{sm:mass_def} m^2(t) = m^2[\phi_0] = m_0^2 + R_2[\phi_0]
\end{equation}
defines the time-dependent in-medium mass. The term in the r.h.s.
of (\ref{sm:mot}) is
\begin{align} 
Q[\phi_0, \phi] &= \ddot\phi_0 + m_0^2 \phi_0 + R_1[\phi_0] +
Q_2[\phi_0,
\phi], \nonumber \\
 \label{sm:qu2} Q_2[\phi_0, \phi] &=
\frac{1}{2} \frac{d R_2[\phi_0]}{d\phi_0}\, \phi^2.
\end{align}
As a result of averaging of Eq. (\ref{sm:mot}), the equation of
motion for the background field is obtained
\begin{equation}
\label{sm:mean} \ddot\phi_0 +m_0^2 \phi_0 + R_1[\phi_0] +
<Q_2[\phi_0, \phi]> = 0,
\end{equation}
where the time independence of the averaging procedure is taken
into account.

The  space-homogeneity assumption implies that the function
$<Q_2[\phi_0, \phi]>$ in Eq. (\ref{sm:mean}) can depend  only  on
time. As it follows from Eqs. (\ref{sm:qu2}) and (\ref{sm:mean}),
the source term in the r.h.s. of Eq. (\ref{sm:mot}) is exclusively
defined by fluctuations,
\begin{equation}\label{sm:qu1}
Q[\phi_0, \phi] = Q_2[\phi_0,\phi] - <Q_2[\phi_0, \phi]>.
\end{equation}
On the other hand, in a non-stationary situation the field
function $\phi(x)$ allows the decomposition:
\begin{eqnarray} \label{sm:dec2} \phi(x) = \frac{1}{\sqrt{V}}\sum_{\mathbf{p}}
\{\phi^{(+)}(\mathbf{p},t) e^{- i \mathbf{p} \mathbf{x}}
 +\phi^{(-)}(\mathbf{p},t) e^{ i \mathbf{p} \mathbf{x}}  \},
\end{eqnarray}
where  $\phi^{(\pm)}(\mathbf{p}, t)$ are the
 positive and negative frequency solutions of the equation of motion
\begin{equation}
\label{sm:12} \ddot\phi^{(\pm)}(\mathbf{p}, t) +
\omega^2(\mathbf{p},t) \phi^{(\pm)}(\mathbf{p}, t) =
 - Q[\phi_0, \phi ; \pm\mathbf{p}]
\end{equation}
with
\begin{equation}\label{sm:omega}
\omega^2(\mathbf{p},t) = m^2(t) +\mathbf{p}^2
\end{equation}
and  with the Fourier image $Q[\phi_0, \phi ; \mathbf{p}]$ of the
 function $Q[\phi_0, \phi]$,
\begin{equation}
Q[\phi_0, \phi] = \frac{1}{\sqrt{V}}\sum_{\mathbf{p}} Q[\phi_0,
\phi ; \mathbf{p}] e^{- i \mathbf{p} \mathbf{x}} .
\end{equation}
The function $Q[\phi_0, \phi ; \mathbf{p}]$ gives a nonlinear
contribution to Eq. (\ref{sm:12}). We suppose that there exists a
finite limit $\lim\limits_{t\to
-\infty}\phi^{(\pm)}(\mathbf{p},t)= \phi_-^{(\pm)}(\mathbf{p})$ in
the infinite past and  assume that solutions
$\phi^{(\pm)}(\mathbf{p},t)$ become asymptotically free
$\phi^{(\pm)}(\mathbf{p},t)\to e^{\pm i\omega_- t}$, where
$\omega_-(\mathbf{p})=\lim\limits_{t\to
-\infty}\omega(\mathbf{p},t)$. The existence of the last limit is
based on the adiabatic hypothesis about switching off the
self-interaction in Eq.(\ref{sm:mass_def}).

After the decompositions (\ref{sm:back}) and (\ref{sm:dec}) the
Hamiltonian density is
\begin{equation}
\label{sm:15} H[\Phi] = H[\phi_0] + H_1[\phi_0, \phi] +
H_2[\phi_0, \phi] +V_r[\phi_0, \phi],
\end{equation}
where $H_0[\phi_0]$ is the background field Hamiltonian, and he
terms $H_1$ and $H_2$ are the Hamiltonian functions of the first
and second order with respect to the fluctuation field
\begin{align}
\label{sm:16} &H[\phi_0] = H_0[\phi_0]+V[\phi_0] = \frac{1}{2}
\dot\phi^2_0 +
\frac{1}{2} m_0^2 \phi_0^2 + V[\phi_0], \\
\label{sm:17} &H_1[\phi_0, \phi] =  \dot \phi_0 \dot \phi + (m^2_0
\phi_0 + R_1[\phi_0]) \phi,\\
\label{sm:18} &H_2[\phi_0, \phi] =  \frac{1}{2} \dot \phi^2 +
\frac{1}{2}(\nabla \phi)^2 +  \frac{1}{2} m^2 \phi^2.
\end{align}
The quasiparticle representation for fluctuations is constructed
by means of the decompositions (\ref{8}). In order to obtain
equation of motion for operator $a(\mathbf{p}, t)$, let us write
the corresponding action with the Hamiltonian (\ref{sm:16}) --
(\ref{sm:18})  (this way is alternative to the Hamiltonian
approach of Sect. \ref{2.1})
\begin{eqnarray} \label{sm:24}
S[\phi ] = \int d^4x\ \{\pi \dot \phi - H_1-H_2 - V_r\}.
\end{eqnarray}
If decompositions (\ref{8}) are substituted here, we { get}
\begin{align}  S[\phi ]& = \frac{1}{2}\sum_{\mathbf{p}} \int dt\  \{
i [a^{\dagger}(\mathbf{p}, t)\dot{a}(\mathbf{p}, t) -
a(\mathbf{p}, t)\dot{a}^{\dagger}(\mathbf{p}, t)] \nonumber\\ &-
\frac{\dot \omega(\mathbf{p}, t)}{\omega(\mathbf{p}, t)}\left[
a^{\dagger}(\mathbf{p}, t) a^{\dagger}(-\mathbf{p}, t) -
a(-\mathbf{p}, t) a^{\dagger}(\mathbf{p}, t)\right] \label{sm:25}
\\ &- \omega(\mathbf{p}, t)[a^{\dagger}(\mathbf{p}, t) a(\mathbf{p}, t)
+ a(\mathbf{p}, t) a^{\dagger}(\mathbf{p}, t)] - 2 V_{r\mathbf{p}}
[\phi_0, \phi] \} + S_1[\phi],\nonumber
\end{align} 
where $S_1[\phi]$ is the part of the action corresponding to the
Hamiltonian (\ref{sm:17}) and $V_{\mathbf{p}}[\phi_0, \phi]$ is
the Fourier image of the { residual} potential term. Variation
with respect to $a(\mathbf{p}, t)$ and subsequent transition to
the occupation number representation lead to the Heisenberg-type
equations of motion  $(\mathbf{p} \ne 0)$
\begin{eqnarray}
\label{sm:26} \dot a(\mathbf{p}, t) = \frac12 W(\mathbf{p}, t )
a^{\dagger}(-\mathbf{p}, t) - i [H_2 + V_r, a(\mathbf{p}, t)],
\end{eqnarray}
where
\begin{eqnarray} \label{sm:27}
W(\mathbf{p}, t) = \frac{\dot \omega (\mathbf{p},t)} {\omega
(\mathbf{p}, t)}=\frac{\dot R_2[\phi_0]} {2\omega^2}.
\end{eqnarray}
In Eq. (\ref{sm:26}) the condensate contribution generated by the
action part $S_1[\phi]$ is omitted because it corresponds to
$\mathbf{p} = 0$ (an appropriate mechanism of the condensate state
with $\mathbf{p} = 0$ and excitations is absent in the present
model).

Let us introduce the distribution function of quasiparticles,
according to Eq. (\ref{distr_func}).  Using methods of Sect.
\ref{SF} and Eq. (\ref{sm:26}) we get the kinetic equation in the
nondissipative approximation, $V_r[\phi_0, \phi] \to 0$,
\begin{eqnarray} \label{sm:ke}
\dot{f} = \frac{1}{2} W u, \qquad \dot{u} = W(1 + 2f) - 2\omega
v,\qquad \dot{v} = 2 \omega u ,
\end{eqnarray}
 which is an analog of (\ref{ODE_SYS}) with the replace of $\Delta$
by $W$. To rewrite Eq. (\ref{sm:mean}) for the background field in
the non-dissipative approximation, one has to calculate the
averaged value $<in \mid \phi^2(x) \mid in>$. In the
space-homogeneous case one can  obtain
\begin{eqnarray}
\label{sm:33} <in \mid \phi^2(x) \mid in> = \frac{1}{2} \int
\frac{d\mathbf{p}} {\omega(\mathbf{p}, t)}[1+2f(\mathbf{p}, t)+
u(\mathbf{p},t)],
\end{eqnarray}
Then Eq. (\ref{sm:mean}) is reduced to
\begin{eqnarray}
\label{sm:34}\label{sm:master}
 \ddot \phi_0 + m^2_0 \phi_0 + R_1[\phi_0] + \frac{1}{2}
\frac{dR_2}{d\phi_0}
 \int \frac{d\mathbf{p}}{\omega(\mathbf{p}, t)} \left[f(\mathbf{p},t)
 +\frac12 u(\mathbf{p},t)\right] = 0
\end{eqnarray}
(the vacuum term is omitted  in integrand).

The KEs (\ref{sm:ke})  and  (\ref{sm:34}) form the closed set of
nonlinear equations describing the back-reaction problem. In the
case of $v[\Phi_0, \Phi] = 0$, this set of equations directly
follows from nonperturbative  dynamics and assumption
(\ref{sm:back}). For the description of the particle creation we
will use the particle density (\ref{dens}) as well as the
background field energy $\epsilon_{cl}$ and energy of created
quasiparticles $\epsilon_{q}$
\begin{eqnarray}
  \epsilon_{cl} &=& \frac12 \dot{\phi}_0^2 +\frac12 m_0^2 \phi_0^2 + V(\phi_0), \nonumber \\
  \epsilon_{q} &=& \int \frac{d^3 p}{(2\pi)^3}\omega(\mathbf{p},t) f(\mathbf{p},t).
  \label{sm:energy}
\end{eqnarray}
The conservation of the full  system energy  can be shown
analytically.

The constructed formalism allows the consideration of certain
initial states at the time $t=t_0$: Initial excitation of the
background field $\phi_0(t_0)$ and $\dot\phi_0(t_0)$ is given
under the additional condition that either $f(\mathbf{p},t_0)=0$
or $f(\mathbf{p},t_0)\neq 0$, where $f(\mathbf{p},t_0)$ is some
initial plasma distribution.

\subsection{Examples}\label{meta3}

\subsubsection{$\Phi^4$ potential}\label{6.3.1} The separation of
the background field (\ref{sm:back}) in the potential
\begin{equation}\label{sm:phi4}
V[\Phi]=\frac{1}{4}\lambda \Phi^4, \qquad \lambda >0
\end{equation}
leads to the following decomposition coefficients (\ref{sm:6}):
\begin{align}\label{sm:coef}
R_1[\phi_0]&=\lambda \phi_0^3, &
R_2[\phi_0]&=3\lambda\phi_0^2,\nonumber \\
V[\phi_0]&=\frac{1}{4}\lambda \phi_0^4, &
V_r[\phi_0,\phi]&=\lambda(\phi_0+\phi/4)\phi^3.
\end{align}
Thus, the time-dependent quasiparticle mass of the fluctuating
field (\ref{sm:mass_def}) is equal  to
\begin{equation}\label{sm:mphi4}
m^2(t)=m_0^2+3\lambda\phi_0^2.
\end{equation}
\begin{figure}[t]
\centering
\includegraphics[width=0.48\textwidth,keepaspectratio]{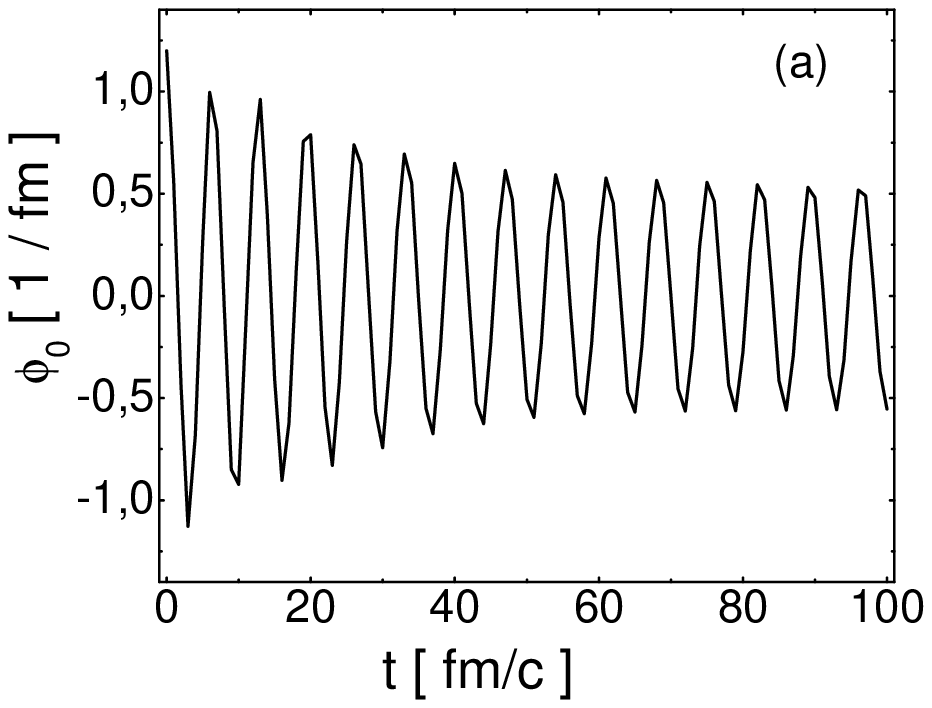}
\includegraphics[width=0.48\textwidth,keepaspectratio]{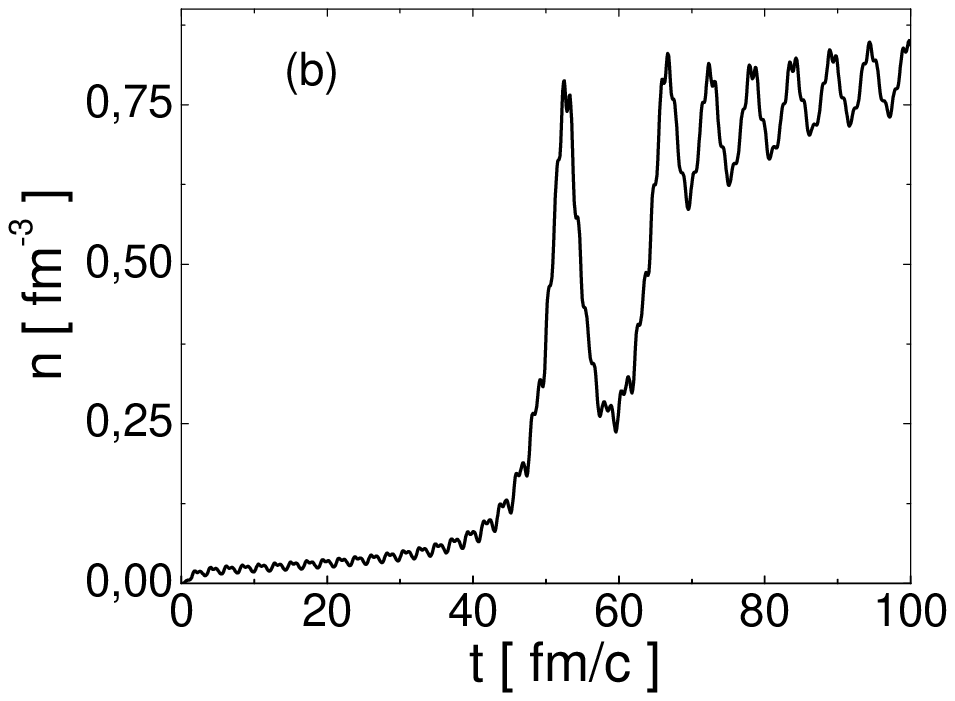}
\includegraphics[width=0.48\textwidth,keepaspectratio]{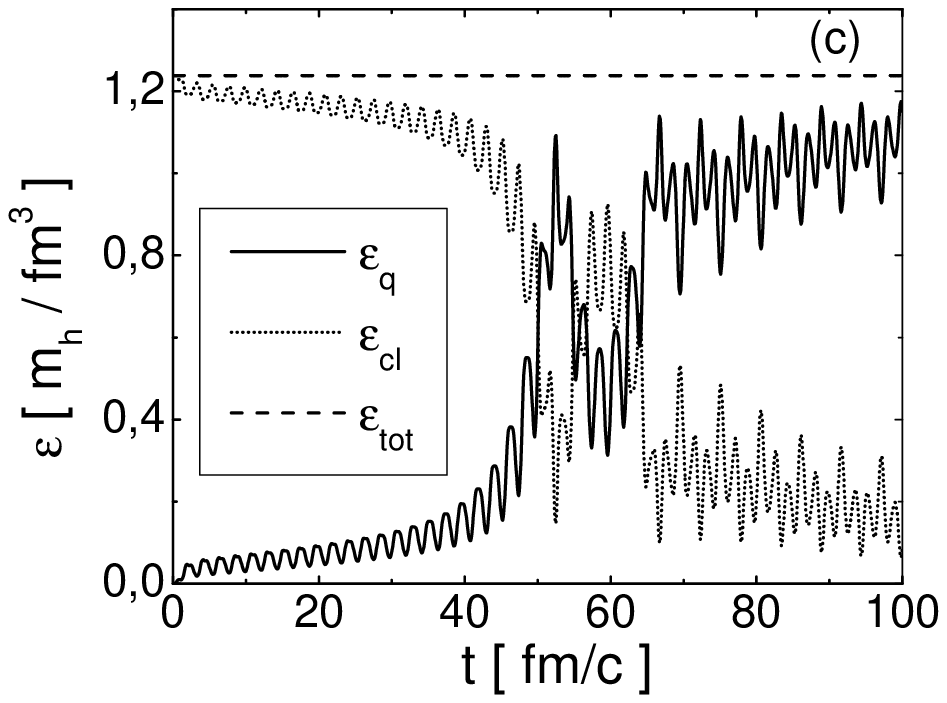}
\includegraphics[width=0.48\textwidth,keepaspectratio]{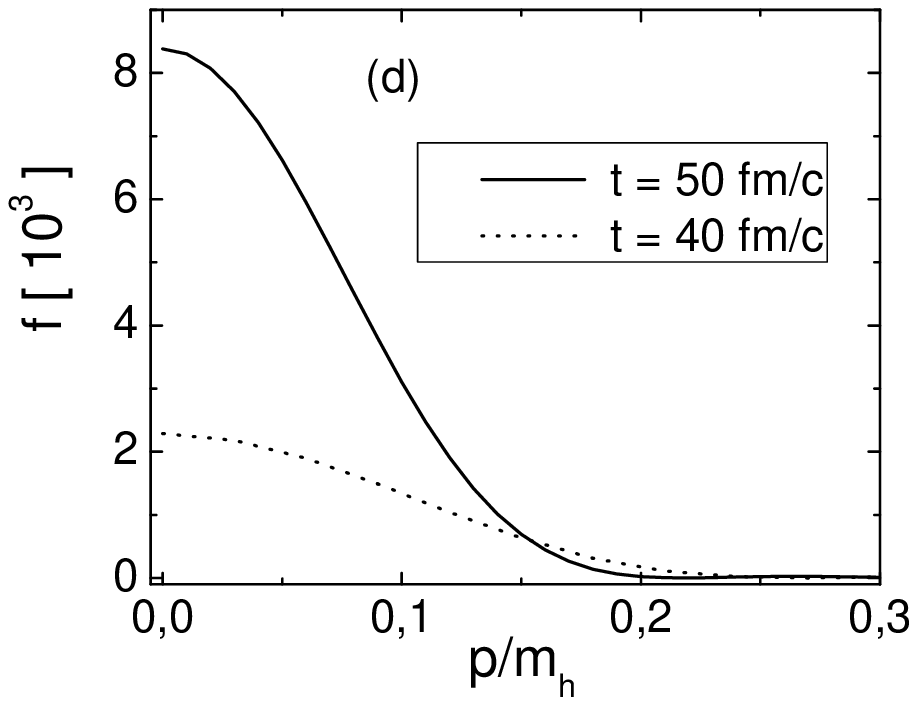}
\caption{Time evolution for the symmetric $\Phi^4$ potential: (a)
the mean field; (b) the particle density; (c) the energy density;
(d) the momentum spectra of particles at time moments $t=40$ fm/c
and $t=50$ fm/c. Parameter values used are: $\lambda=1, \
m_0=m_h=197$ Mev, $\phi_0(0)=1.2 fm^{-1}$, $\ \dot{\phi}_0(0)=0$.
\label{fig18}}
\end{figure}
If $\lambda<0$ and the excitation  is strong enough, it is
possible that   a tachyonic mode will arise that corresponds to an
unstable state \cite{AK}. The mass (\ref{sm:mphi4}) determines the
factor (\ref{sm:27})
\begin{equation}
W(\mathbf{p},t)=\lambda\frac{2\phi_0\dot\phi_0}{\omega^2(\mathbf{p},t)}.
\end{equation}
The KE (\ref{sm:ke}) with this factor is correct in the
nondissipative approximation where the residual potential $V_r$ is
neglected.

Let us write down also the equation of motion for the background
field (\ref{sm:master}) in this approximation:
\begin{equation}\label{sm:317}
\ddot\phi_0+M^2(t)\phi_0+\lambda\phi_0^3=0
\end{equation}
with the corresponding mass to be equal to
\begin{equation}\label{sm:(3.1.8)}
M^2(t)=m_0^2+3\lambda\int\frac{d^3k}{\omega(\mathbf{p},t)}
\left[f(\mathbf{p},t)+\frac12 u(\mathbf{p})\right],
\end{equation}
{\em i.e.} the mass of the condensate excitations is defined by
both the distribution of quasiparticles  and vacuum polarization.

In numerical calculations we apply  zero initial conditions for
the distribution function and nonzero ones for the background
field $\phi_{0}(t_{0})=1.2$ ( here and below we use the
 units $\hbar=c=1$).   Parameter values are chosen by analogy with
  \cite{Baacke}, where the authors offered an alternative method for
describing quantum systems under action of the strong background
field (so-called Cornwall-Jackiw-Tomboulis method \cite{CJT}). As
is seen in Fig.\ref{fig18}, at the early evolution stage all
energy is mainly concentrated in field oscillations. For $t < 50$
the particle number density grows slowly. However, density
drastically increases at $t \sim 50$ and after
this time the quantum energy dominates over classical one.%

The case $\lambda <0$ (absolutely unstable potential) is
associated with a tachyonic regime which is realized for high
enough excitations when the initial amplitude $\phi_0(t_0)$
satisfies the condition $m_0^2+3\lambda \phi_0^2(t_0)\le 0$.

\subsubsection{Double-well potential \label{sm:w}}
The potential
\begin{equation}\label{sm:321}
V[\Phi]=\frac{1}{4}\lambda \Phi^4-\frac{1}{2}\mu^2\Phi^2, \qquad
\lambda >0
\end{equation}
leads to the same equation (\ref{sm:317}) for the background field
but with a new mass (we put here $m_0=0$ and $\mu^2>0$)
\begin{equation}\label{sm:3.2.2}
M^2(t) =
-\mu^2+3\lambda\int\frac{d\mathbf{p}}{\omega(t)}\left[f(\mathbf{p},t)+
\frac12 u(\mathbf{p},t)\right].
\end{equation}
The factor (\ref{sm:27}) equals
\begin{equation}\label{sm:3.2.3}
W(\mathbf{p},t)=\lambda
\frac{3\phi_0\dot\phi_0}{2\omega^2(\mathbf{p},t)},
\end{equation}
where now the quasiparticle frequency (\ref{sm:omega}) includes
the time-dependent mass
\begin{equation}\label{sm:(3.2.4)}
m^2(t)=-\mu^2+3\lambda\phi_0^2.
\end{equation}

The vicinity of the central point $\phi_0(t)=0$ is an  instable
region.  In this region the group velocity $v_g=d\omega(k)/dk
=k/\omega(k)$ is  either super-luminous for $k>k_c$, where $k_c$
is the root of the equation $\omega(k,t)= 0 $ or undefined for $ k
< k_c$. Thus, it is the tachyonic region.
\begin{figure}[t]
\centering
\includegraphics[width=0.48\textwidth,keepaspectratio]{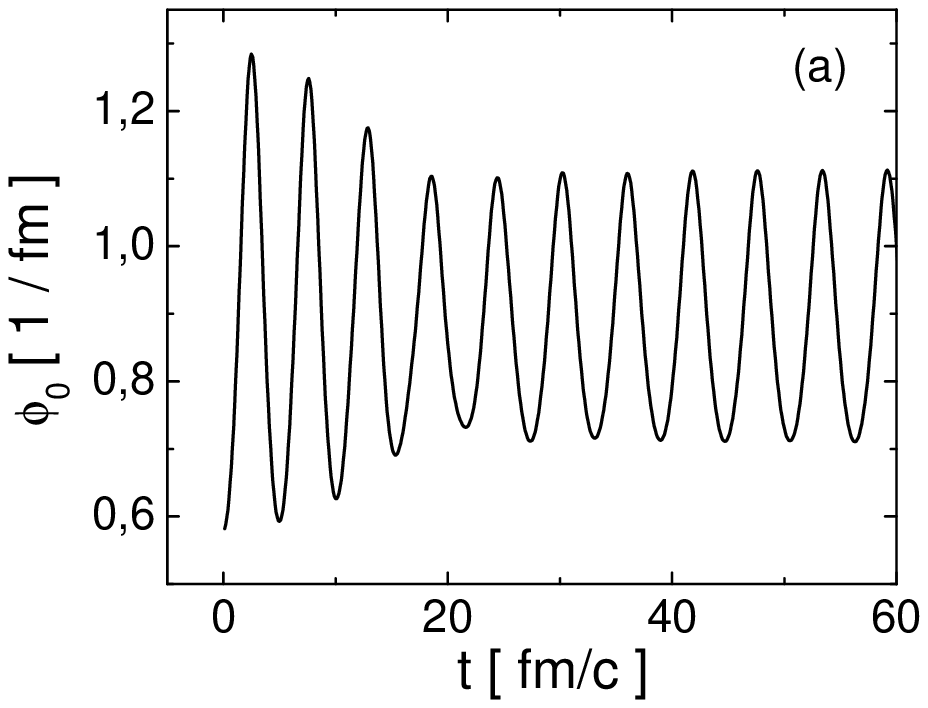}
\includegraphics[width=0.48\textwidth,keepaspectratio]{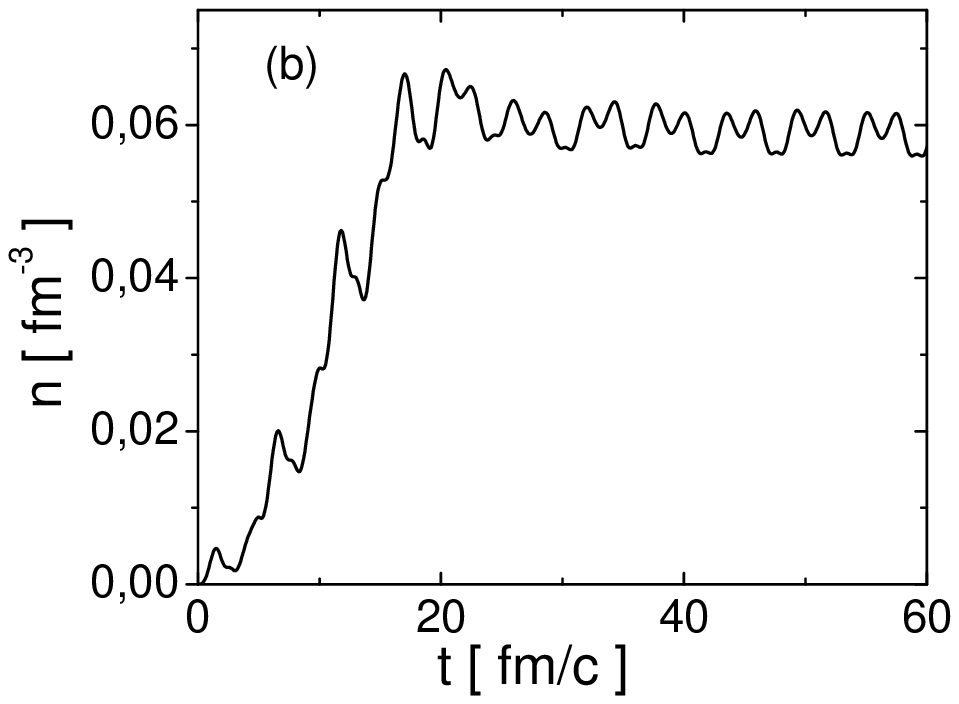}
\includegraphics[width=0.48\textwidth,keepaspectratio]{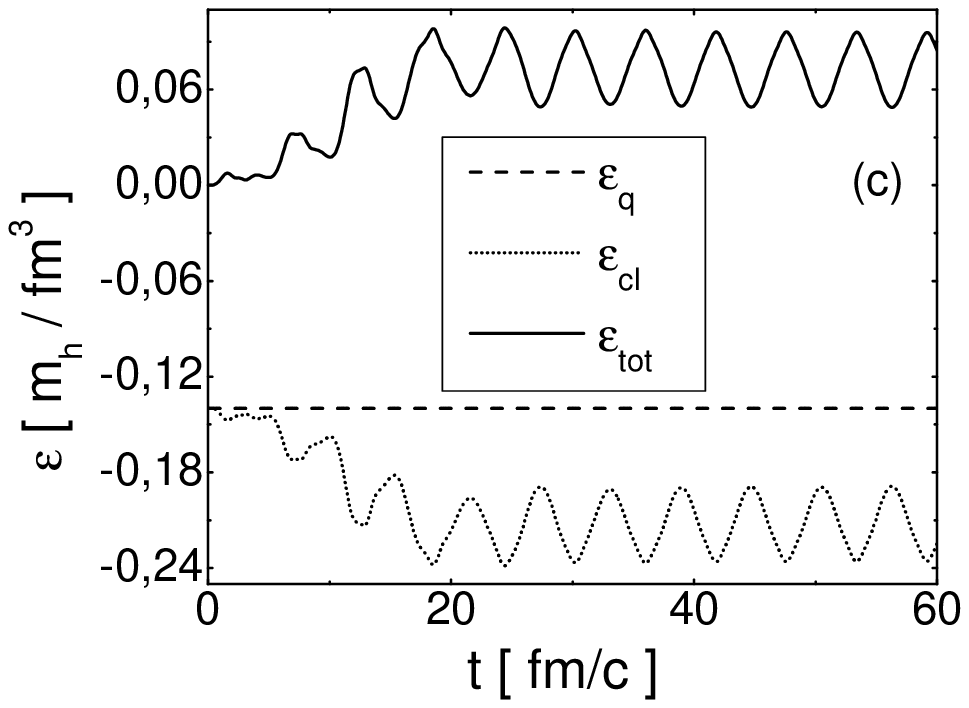}
\includegraphics[width=0.48\textwidth,keepaspectratio]{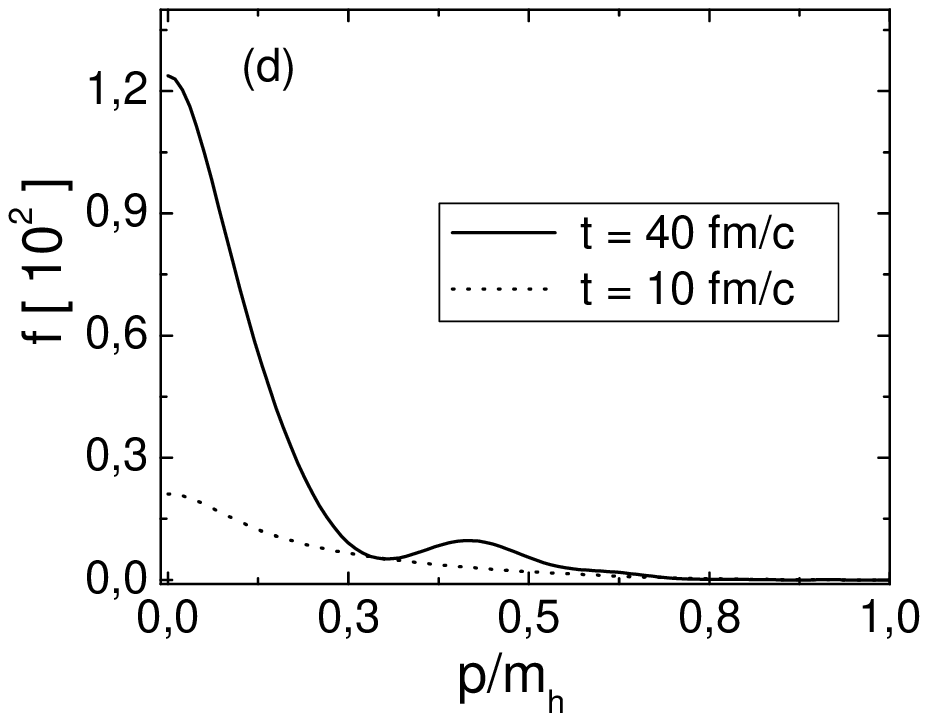}
\caption{Time evolution for the bistable $\Phi^4$ potential
(\ref{sm:321}): (a) the mean field; (b) the particle density; (c)
the energy density; (d) the momentum spectra of particles at time
$t=10$ fm/c and $t=40$ fm/c. Parameter values are $m_0=0,
\mu=m_h$, $\lambda=6$, $\phi_0(0)=0.58 fm^{-1},
\dot{\phi}_0(0)=0$.} \label{fig_19}
\end{figure}

Let us denote the minimum position of the potential (\ref{sm:321})
as $\Psi_{\pm} = \pm\Psi_0 = \pm\mu/\sqrt{\lambda}$ and put the
new origin of the reference frame in one of these points,
$\Phi=\Psi_{\pm}+\Psi$. We discriminate now the background
component $\phi_0$ and the field $\Psi$, {\em i.e.}
$\Psi=\phi_0+\phi$. We will omit the sign indices $(\pm)$, which
identify  the branch of $\Psi_{\pm}$ which  the fields $\Psi ,
\phi_0$ and $\phi$ belong to. Only rather a small range of
excitations in the vicinity of the stable points $\Psi_{\pm}$ will
be considered below, $|\phi (x)|, |\phi_0(t)| \ll \sqrt{2}
\Psi_0$, where $\pm \sqrt{2} \Psi_0$ are the roots of the equation
$V[\Psi] = 0$. Using Eq.(\ref{sm:mot_g}) and methods described in
Sect. \ref{meta2}, one can obtain the following system of
equations of motion:
  \begin{eqnarray}
  \ddot\phi_0+2\mu^2\phi_0+3\lambda\Psi_{\pm}\phi_0^2+\lambda\phi^3_0
  +3\lambda(\Psi_{\pm}+\phi_0)\langle\phi^2\rangle +
  \lambda\langle\phi^3\rangle =0 , \nonumber\\ \label{sm:mot_phi}
-[\partial_\mu\partial^\mu + m^2_{\pm}(t)]\phi+
3\lambda(\Psi_{\pm}+\phi_0)[\langle\phi^2\rangle -\phi^2]+
\lambda[\langle\phi^3\rangle -\phi^3]=0,
  \end{eqnarray}
where
\begin{equation}\label{sm:mphi}
m^2_{\pm}(t)=2\mu^2+3\lambda\phi_0(\phi_0+2\Psi_{\pm}).
\end{equation}
In the nondissipative approximation  only linear terms should be
kept in Eq.(\ref{sm:mot_phi}). This leads to the KE (\ref{sm:ke})
with the factor defined by the time-dependent mass (\ref{sm:mphi})
\begin{equation}\label{sm:amplphi}
  W_{\pm}(\mathbf{p},t)=\frac{\dot\omega(\mathbf{p},t)}{2\omega(\mathbf{p},t)}=
\frac{3\lambda\dot\phi_0(\phi_0+\Psi_{\pm})}{2\omega^2(\mathbf{p},t)}.
\end{equation}

The mean values $\langle\phi^2\rangle$ and $\langle\phi^3\rangle$
are calculated  either in the minimal order of the perturbation
theory (for $\lambda\ll 1$) or in the random phase approximation.
We use the result (\ref{sm:33}) for $\langle\phi^2\rangle$ and
$\langle\phi^3\rangle =0$. Thus, we have
\begin{multline}\label{sm:329}
\ddot\phi_0+ \lambda\phi_0 \biggl[(2+3\phi_0)\Psi_{\pm}+\phi_0^2
\biggr] \\ + 3\lambda
(\Psi_{\pm}+\phi_0)\int\frac{d\mathbf{p}}{2\,\omega(\mathbf{p},t)}\left[2f(\mathbf{p},t)+
v(\mathbf{p},t)\right] =0.
\end{multline}
Numerical results for the  set of parameters corresponding to
\cite{Baacke} are presented in Fig.~\ref{fig_19}. The stationary
regime is achieved faster than in the case of the symmetric
potential.

Another formalism for describing the  strong field problem in the
quantum field system with the potential (\ref{sm:321}) is
developed by J. Baacke {\em et al} (see \cite{Baacke}, and papers
cited therein).  Here the general kinetic approach is developed
for arbitrary highly-excited nonequilibrium states in the scalar
QFT with self-interaction admitting the existence of unstable
vacuum states. We restrict ourselves to the collisionless
(non-dissipative) approximation. However, attempts to go beyond
this approximation have been made \cite{CI}. As a particular
example, $\phi^4$ and double-well potentials were investigated.

It would be of interest to study some other properties of the
considered model, such  features as  excitation transitions
between states with different vacua (in the same space-time
point). Apparently, it is possible  at high initial excitation
$|\phi_0 (0)| \ge \sqrt{2} \Psi (0)$ and  as a consequence of the
tunnelling process through the central barrier. The last problem
is especially interesting in the generalized double-well potential
model with non-degenerated vacuum states.

\subsubsection{$\eta$-meson system}

We will use now the developed technique to consider a more
complicated quantum field system, namely, the $\eta$-meson sector
of the Witten-DiVecchia-Veneziano model \cite{Veneziano} which
describes the low-energy dynamics of the nonet of pseudoscalar
mesons in the large $N_c$ limit of QCD. We drow our attention to
the singlet state of this model with the following Lagrangian
\cite{Ahrensmeier}:
\begin{equation}\label{4.2}
{\cal L}= \frac{1}{2} {\partial}_{\mu} {\eta}{\partial}^{\mu}
{\eta} + b^2 {\mu}^2 \cos\Big(\frac{\eta}{b}\Big) - \frac{a}{2}
{\eta}^2 ,
\end{equation}
where $b=\sqrt{{3}/{2}} b_{\pi}$, $b_{\pi}=92$ MeV is the
semi-leptonic pion decay constant, ${\mu}^2= \frac{1}{3}
(m^2_{\pi} + 2m^2_{K})\simeq 0.171~ {\rm GeV}^2$, $m_\pi$ and
$m_K$ are $\pi$- and $K$-meson masses,
 $a = m^2_{\eta} + m^2_{{\eta}^{\prime}} -2 m_K^2 \simeq 0.726~ {\rm GeV}^2$
for zero temperature. The corresponding total Hamiltonian density
is given by
\begin{eqnarray}
H = \frac12\dot{\eta}^2 + \frac12 (\nabla\eta)^2 +\frac12 a\eta^2
+ 2 b^2\mu^2 \sin^2{\frac{\eta}{2b}},\label{4.4}
\end{eqnarray}
where the constant addend $b^2\mu^2$ was discarded.

The Hamiltonian density (\ref{4.4}) can be reduced to the form
\cite{Veneziano}
\begin{eqnarray}
&&H = \frac12\dot{\eta}^2 + \frac12 (\nabla\eta)^2 +\frac12
m^2_0\eta^2+ H_{in},\nonumber \\ &&H_{in} =2 b^2\mu^2
\left[\sin^2{\frac{\eta}{2b}}
-\left(\frac{\eta}{2b}\right)^2\right] ,\label{4.6}
\end{eqnarray}
where $m_0^2=a+\mu^2$. The effective potential $H_{in}$ is
constructed in such a way that its formal decomposition with
respect to the field function $\eta(x)$ does not contain the
corresponding squared contribution to be associated with the mass
term. However, the redefinition (\ref{4.6}) leads to the
absolutely unstable potential (Sec. \ref{6.3.1}) with the
corresponding tachyonic modes \cite{Veneziano}. In our opinion,
these modes have artificial character and can be eliminated by
return to the original Hamiltonian density (\ref{4.4}).

\begin{figure}[t]
\centering
\includegraphics[width=0.48\textwidth,keepaspectratio]{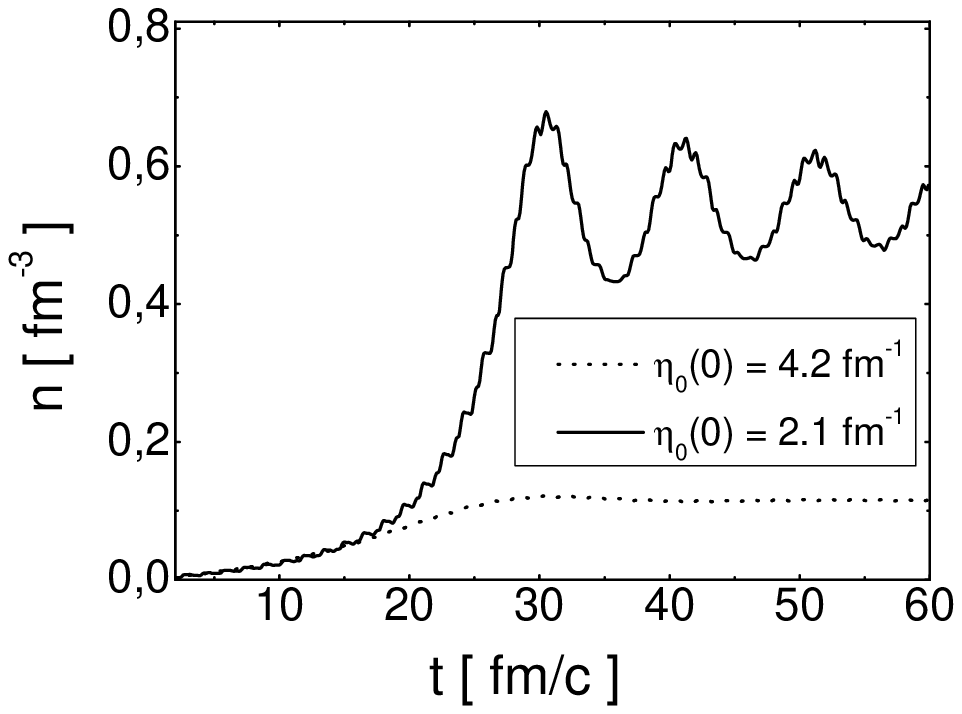}
\includegraphics[width=0.48\textwidth,keepaspectratio]{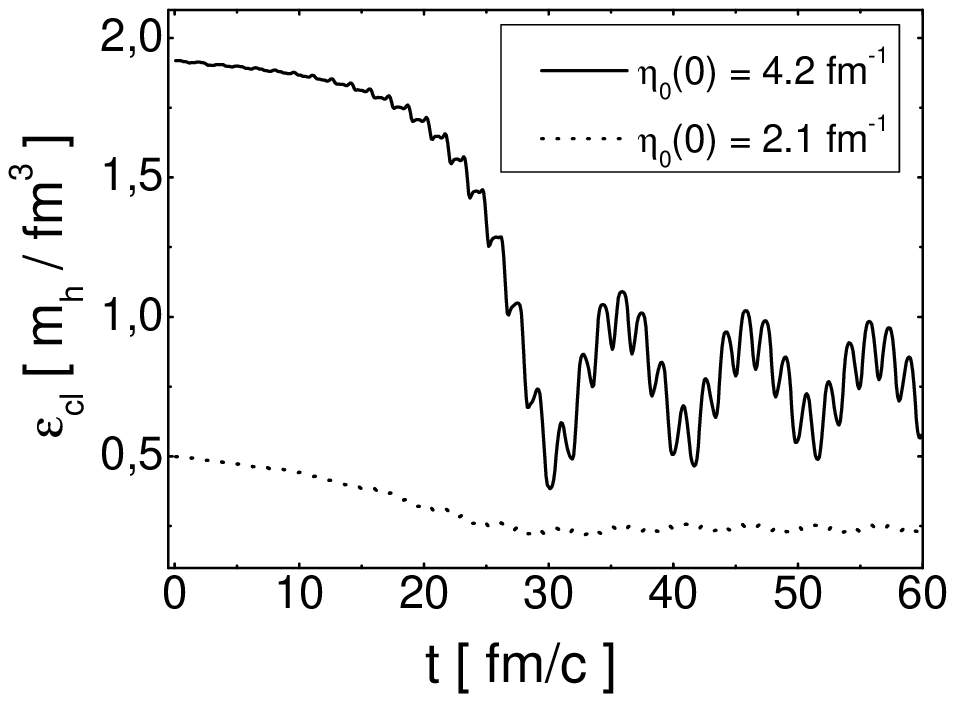}
\caption{The time evolution for the potential (\ref{4.6}):  The
particle density (left panel) and the mean field energy (right
panel). Parameter values are $m_0=0$, $\dot{\eta}_0(0)=0$.
\label{fig_20}}
\end{figure}

By analogy with Eq. (\ref{sm:back}), let us select the
quasiclassical field $\eta_0(t)$ and quantum fluctuation part
$\phi(x)$,
\begin{equation}\label{4.7}
  \eta(x) = \eta_0(t)+\phi(x).
\end{equation}
Now we can implement the general formalism of Sec. \ref{sm:w} to
the $\eta$-meson system with the Hamiltonian density (\ref{4.4}).
The master equation (\ref{sm:master}) is now given by
\begin{eqnarray}\label{4.8}
 \ddot \eta_0 + a \eta_0 + b \mu^2 \sin{\left(\frac{\eta_0}{b}\right)}
\left[1 - \frac{1}{2b^2}\int \frac{d\mathbf{p}}{\omega} \left(f+
 \frac12 v\right)\right] = 0
\end{eqnarray}
where $\omega(\mathbf{p},t)$ is the quasiparticle energy
(\ref{sm:omega}) with the mass (\ref{sm:mass_def})
\begin{equation}\label{4.9}
  m^2(t) = a+\mu^2 \cos{\frac{\eta_0}{b}}.
\end{equation}
The KE (\ref{sm:ke}) is defined via the factor (\ref{sm:27}) which
can be presented as
\begin{equation}\label{4.10}
  W(\mathbf{p},t)= -\frac{\mu^2}{4b\,\omega^2(\mathbf{p},t)}\,\dot{\eta}_0\,
  \sin{\frac{\eta_0}{b}}.
\end{equation}
The mass formula (\ref{4.9}) allows one to fit parameters $a$,
$\mu^2$ and $m^2(t)>0$ for any amplitude of the quasiclassical
field $\eta_0(t)$, as is seen in Fig. \ref{fig_20}. The
qualitative behavior of the presented observable is similar to
that for the bistable potential case. Oscillations of the
condensate energy are more pronounced for the symmetric potential
that is related with higher initial values of $\eta_0$.

 It was shown that if one vacuum state among a set of vacuum
states is fixed, it leads to some specific evolution of the system
under action of the mean field. But there is an open question
concerning occupation of other system states due to tunnelling
process \cite{Linde}. This problem follows also from our
treatment.

\section{Summary \label{SUM}}

This review is mainly based on the results obtained by the authors
in the last years and aims to summarize the information about the
kinetic description of vacuum particle creation. The latter
results for the inertial mechanism with the time-dependent
particle masses are stipulated at the phenomenological level.
Three basic quantum field models were considered here: Massive
scalar, vector and spinor fields. The constructed kinetic theory
was applied (Sect. \ref{UNI})  to a conformal cosmology model for
investigation of matter created from vacuum in an early period of
the Universe evolution. In particular, it was shown that the
density of the produced vector bosons is sufficient for
explanation of the present-day density of CMB photons.

The obtained results can be used  for a subsequent study of
different aspects of matter dynamics  created from the vacuum in
the early Universe (the equation of state, the long wave-length
acoustic excitations, the back-reaction problem etc.).  Only the
single mechanism of a mass change was considered where this change
is induced by the conformal expansion of the Universe whose action
is switched on for the mass $m_0 (t_0)$ at some arbitrary initial
time $t_0$. For construction of a more elaborated theory, one
should eventually take into account the inflation mechanism  of
mass generation acting during an earlier period of the Universe
evolution \cite{Dolgov,22,24,25}. It would be of interest to
consider the generation of particles of different masses and
quantum statistics using Eq. (\ref{mass}) which is valid for all
particles independently of their inner symmetry.

Interesting perspectives are opened up also by results of Sect.
\ref{meta}, where some simplest quantum field models are
investigated for scalar fields with various self-interactions and
a corresponding quasiclassical nonstationary field (the problem of
a phase transition at the restoration of broken symmetry, particle
tunnelling between states with a different vacua {\em etc.}).

Certainly, the considered examples do not exhaust  all variety of
systems, where the vacuum generation of particles is induced by
the inertial mechanism. In particular, a large class of
experimentally controlled models of meson and quark subsystems
evolved close to a phase transition has remained beyond our scope.
In addition, the investigation of two-particle correlations is of
interest for description of some delicate experimental effects in
nuclear reactions \cite{Andreev96}. The extension of the
mean-field approximation to take into account the back-reaction
and collisions is studied intensively with different methods in
some scalar models \cite{Ahrensmeier,Baacke}. The important
problem of relevant initial conditions also attracts attention in
the context of heavy-ion collisions \cite{Berges}.  We did not
worry much concerning relativistic invariant form of the developed
formalism : it is an open question.The basis for its solution is
the covariant Hamiltonian formalism in relativistic kinetic theory
\cite{Maino,ZZ97,BPZZ06}.

Finally, some effects discussed here take place also in condensed
matter physics and can be interpreted in the conceptual framework
of the inertial mechanism ({\em e.g.}, \cite{Green,Oka} and
references therein). Such a kind of analogy is very promising for
an experimental test of the given theory with more accessible
realizations under condensed matter conditions.

\subsection*{Acknowledgement}
We are grateful to Profs. D.B.~Blaschke and M.P.~D\c{a}browski for
discussion of some cosmological aspects of the present work. We
are thankful to E.E.~Kolomeitsev and V.N.~Pervushin for  careful
reading of the manuscript and valuable remarks.

This work was supported in part by RFBR Grant N
05-02-17695 and by a special program of the Ministry of Education
and Science of the Russian Federation (grant RNP.2.1.1.5409).

\end{document}